\documentclass[pra,aps,twocolumn,10pt,superscriptaddress,notitlepage,longbibliography]{revtex4-2}
\usepackage[dvipsnames]{xcolor}
\usepackage[english]{babel} 
\usepackage{graphicx}
\usepackage{float}
\usepackage{braket}
\usepackage{epstopdf}
\usepackage{mathtools}
\usepackage{amsmath}
\usepackage{amssymb}
\usepackage{bbm,bm}
\usepackage[caption=false]{subfig}
\usepackage{pythonhighlight}
\usepackage{hyperref}
\usepackage{float}
\usepackage{bbold}
\usepackage[T1]{fontenc}

\usepackage[markup=underlined]{changes}
\makeatletter
\@namedef{Changes@AuthorColor}{magenta}
\colorlet{Changes@Color}{magenta}
\makeatother
\usepackage{hyperref}
 \hypersetup{
     colorlinks=true,
     linkcolor=blue,
     filecolor=blue,
     citecolor = magenta,      
     urlcolor=red,
     }

\usepackage{braket}

\usepackage{xargs}
\newcommandx{\greencom}[2][1=]
{\todo[inline, color=green!40,#1]{#2}}
\newcommandx{\bluecom}[2][1=]
{\todo[inline, color=blue!40,#1]{#2}}
\newcommandx{\bluemargin}[2][1=]
{\todo[color=blue!40,#1]{#2}}

\usepackage{letltxmacro}
\LetLtxMacro{\ORIGselectlanguage}{\selectlanguage}
\makeatletter
\DeclareRobustCommand{\selectlanguage}[1]{%
  \@ifundefined{alias@\string#1}
    {\ORIGselectlanguage{#1}}
    {\begingroup\edef\x{\endgroup
       \noexpand\ORIGselectlanguage{\@nameuse{alias@#1}}}\x}%
}
\newcommand{\definelanguagealias}[2]{%
  \@namedef{alias@#1}{#2}%
}

\makeatother

\definelanguagealias{en}{english}
\definelanguagealias{EN}{english}
\definelanguagealias{eng}{english}
\definelanguagealias{de}{ngerman}

\graphicspath{{paper_fig_V2/}} 

\begin{document}

\title{Probing dressed states and quantum nonlinearities in a strongly coupled 
three-qubit waveguide system under optical pumping}

\author{Sofia Arranz Regidor}
\email{18sar4@queensu.ca}
\affiliation{Department of Physics,
Engineering Physics and Astronomy, Queen's University, Kingston, Ontario, Canada, K7L 3N6}
\author{Stephen Hughes}
\affiliation{Department of Physics,
Engineering Physics and Astronomy, Queen's University, Kingston, Ontario, Canada, K7L 3N6}
\email{shughes@queensu.ca}

\date{\today}

\begin{abstract} 
We  study a three-qubit waveguide system in the presence of optical pumping,
when the side qubits act as atomlike mirrors, manifesting in a strong light-matter coupling regime.
The qubits are modelled as two-level systems, where we account for important saturation effects and quantum nonlinearities.
Optically pumping this system is shown to lead to a rich manifold of dressed states that can be seen in the emitted spectrum, and we show two different theoretical solutions using a medium-dependent
master equation model in the Markovian limit, as well as using matrix product states without invoking any Markov approximations.
We demonstrate how a rich nonlinear spectrum is obtained by varying the relative decay rates of the mirror qubits as well as their spatial separation, and show the limitations of using a Markovian master equation. Our model allows one to directly model 
giant-atom phenomena, including important retardation effects
and multi-photon nonlinearities.
We also show how the excited three qubit system,
in a strong coupling regime, deviates  significantly from a Jaynes-Cummings model when entering the nonlinear regime.
\end{abstract}

\maketitle

\section{Introduction}
Waveguide quantum electrodynamics (QED) is important in the study of light-matter interactions in quantum optical circuits~\cite{RevModPhys.93.025005}, allowing a controlled coupling between two-level systems (TLSs) or quantum bits (qubits) and a continuum of quantized modes \cite{PhysRevA.76.062709,RevModPhys.89.021001}. Waveguide QED systems give rise to new effects not observed in free space quantum optics or in traditional cavity-QED \cite{RevModPhys.95.015002}. In particular,
the quasi-one-dimensional (1D) confinement enhances the qubit coupling, allowing one to manipulate light-matter interactions between qubits and waveguide mode photons \cite{PhysRevA.102.023702,PhysRevResearch.2.043213}.
Moreover, an ensemble of qubits in a waveguide generates strongly correlated photon transport beyond the dipole-dipole interaction regime \cite{PhysRevLett.121.143601}, allowing one to study rich many-body dynamics.

Although waveguide-QED systems are excellent systems for confining photons to waveguide modes, they naturally dissipate, which can make experimental demonstrations difficult, e.g., for confining photons and realizing nonlinear resonances.
Different approaches have been investigated in order to overcome this limitation, with the utilization of so-called ``giant atoms'' being one of the most successful proposals~\cite{Liu:22,PhysRevLett.128.223602,PhysRevA.106.013702,Wang:21}. A giant atom configuration can enable a decoherence-free interaction with waveguide-QED \cite{PhysRevLett.120.140404,Kannan2020,PhysRevResearch.2.043184}, by manipulating the phase between different qubits in the waveguide.
Various material systems can realize waveguide QED implementations, including superconducting circuits
\cite{Mirhosseini2019,Kannan2020,Masson2022}
and quantum dots \cite{PhysRevX.2.011014,PhysRevB.75.205437,Lund-Hansen2008}.
Another unique feature of waveguide QED is the ability to realize and exploit non-Markovian dynamics, which can be realized with time-delayed coherent feedback and with suitably long delay times between multiple qubits
\cite{PhysRevA.95.053807,PhysRevA.101.023807,PhysRevLett.124.043603,PhysRevResearch.3.023030,PhysRevLett.110.013601,PhysRevLett.115.060402,Whalen2017,PhysRevLett.122.073601,PhysRevA.104.053701,PhysRevA.106.013714,PhysRevA.106.023708,PhysRevA.102.043718}.

\begin{figure} [t]
\centering
\includegraphics[width=\columnwidth]{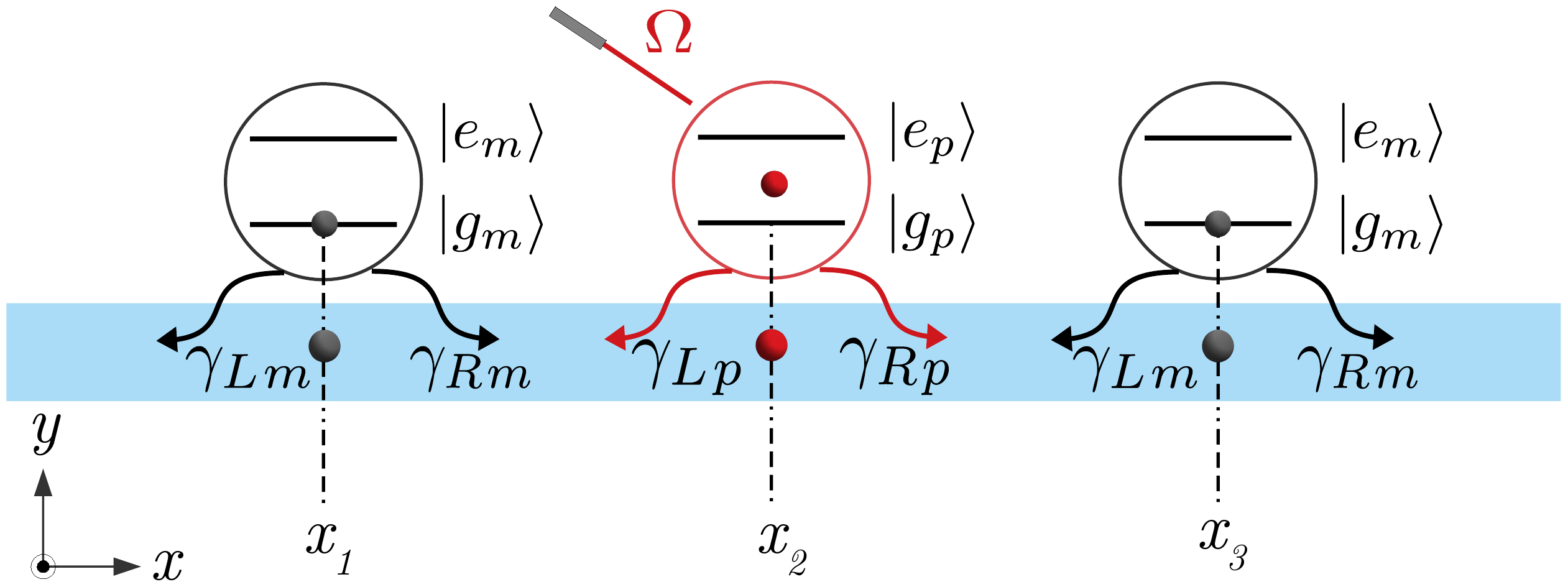}%
 \caption{Schematic of three qubits in a waveguide, with optical pumping of the middle (probe) qubit. }
 \label{schematic}
\end{figure}

One of the defining features of QED systems is for exploring unique quantum nonlinearities, that have no classical analogue,
an example of which is found in the anharmonicity of a driven Jaynes-Cummings (JC) ladder system 
 \cite{Bishop2008,Fink2008,Schuster2008,Illes2015}.
 At the linear response level, vacuum Rabi oscillations can occur~\cite{Agarwal:85}, which can also be explained classically or semiclassically \cite{PhysRevLett.64.2499}. Nevertheless,
 vacuum Rabi splitting is an important prerequisite for exploring unique quantum nonlinearities in the strong coupling regime.

Recently, it was experimentally demonstrated how 
atomlike mirrors \cite{Chang2012,PhysRevA.105.033705}  can realize a strong light-matter coupling regime similar to a JC system, using superconducting qubits \cite{Mirhosseini2019}.
These works focused on the linear response with qubit spatial separations that are in a Markovian regime, and it is interesting to explore how such a system behaves in a nonlinear regime (which is precisely where one may find unique quantum phenomena), which challenges many of the usual quantum optics models.
Namely, how does such a finite-size
cavitylike system responds when optically pumped
to a nonlinear regime, and does the system follow
a (dissipative) JC model or an extended system of three coupled qubits?
In this paper, we directly address this question by modelling an optically pumped target qubit embedded in atomlike mirrors (see Fig. 1). We explore both Markovian and non-Markovian regimes, and show a host of new resonances in the nonlinear regime.

The rest of our paper is organized as follows. 
In Sec.~\ref{sec:theory}, we introduce the main theoretical approaches for modelling the three-qubit waveguide system.
We first present a Markovian master equation solution, and connect it to the medium-dependent Green functions to explain the various decay rates and
photonic coupling effects in a self-consistent way, using a macroscopic QED approach. For the main observable of interest, we use the waveguide-emitted spectrum, which contains contributions from all qubits. Second, we present a matrix product state (MPS) approach,
which allows one to include multiple waveguide photons and fully describe non-Markovian effects, which is important when optical delays are introduced (e.g., for qubits with larger separations, when a Markov approximation is no longer valid).
Third, for reference,
we also show the solution for linear response, previously studied
in Ref.~\cite{PhysRevA.104.L031701}.

In Sec.~\ref{sec:results}, we show various results of the two theoretical models.
We first highlight the linear regime, where this system also gives an analogue of vacuum Rabi splitting (two polariton peaks), which can be modified further for finite separations between the qubits (when the effects of retardation become important \cite{PhysRevA.104.L031701}). 
We  study the  emitted spectrum as a function of the different qubit decay rates, and show how various sharp resonances appear when the decay rates of the mirrors ($\gamma_m$) are greater than those of the probe qubit ($\gamma_p$).
We explain the main spectral peaks, by studying the transitions between the quasi-energy levels (dressed states) from the system Hamiltonian, including the effect of the drive. We subsequently explore how the spectrum changes as a function of pump strength, those features are well explained in terms of allowed transitions between dressed states. We compare these findings with the solutions of a driven JC system and demonstrate how the nonlinear features
of the three qubits emerge and survive even for large pump strengths, showing features that are significantly different from the driven JC model. This is in contrast to the linear response.
Next, we study the waveguide QED system using MPS and investigate the role of retardation, and highlight new nonlinear resonance energies for larger qubit separations. Our conclusions are presented in Sec.~\ref{sec:conclusions}.

\section{Theory}
\label{sec:theory}

In this section, we describe two different approaches to model the dynamics of the multi-qubit waveguide system. First, we use a 
Markovian master equation approach that is derived from a macroscopic Green function formalism ~\cite{Dung2002,PhysRevA.91.051803,PhysRevA.12.1475,Agarwal2013}. 
For a recent review of waveguide QED, 
including such theory techniques, see
Ref.~\cite{RevModPhys.95.015002}.
As discussed in Ref.~\cite{PhysRevA.104.L031701}, these macroscopic approaches fully recover
model Hamiltonian formalisms for waveguide QED in the appropriate limit~\cite{PhysRevA.95.053807,Chang2012}; this is complemented with an approach to obtain the spectrum emitted into the waveguide. Second, we use an MPS approach~\cite{cajitas,PhysRevResearch.2.013238}. The advantage of the latter model is that it does not rely on the Markov approximation, and can model multi-photon states in the waveguide.
The MPS approach also allows us to model the effects of retardation, which is known to be important for longer qubit separations and delay times.
Indeed, longer delay times are known to tune and improve the strong coupling regime~\cite{PhysRevA.104.L031701} at the vacuum level, and below we will investigate what happens in an optical pumping regime beyond weak excitation.
For reference, we also show a frequency-dependent solution for linear response~\cite{PhysRevA.104.L031701}.

\subsection{Markovian master equation using the waveguide Green functions}
\label{subsection:ME}

The photonic Green function for the waveguide medium has the general analytic form~\cite{PhysRevB.75.205437,PhysRevA.104.L031701}
\begin{align}
    {\bf G} &\equiv {\bf} {\bf G}_W({\bf r}_a,{\bf r}_b,\omega) \nonumber \\
    &= iA[ {\bf f}_k({\bf r}_a) {\bf f}_k^*({\bf r}_b) H(x_a-x_b) e^{ik(x_a-x_{b})} \nonumber \\
    &+ {\bf f}^*_k({\bf r}_a) {\bf f}_k({\bf r}_b)     H(x_b-x_a) e^{ik(x_b-x_{a})}],
\end{align}
where $A$ is a constant,
${\bf f}_k({\bf r})$ is the waveguide mode of interest,
and 
the Heaviside step functions ($H$);
when $x_a=x_b$,  $H(0)=0.5$, and the dipole is coupled to both waveguide mode directions (unless, e.g., in a chiral system~\cite{PhysRevLett.115.153901,Sllner2015}).
Note that ${\bf G}$ is a dyad, formed by the outer product of two vectors, but in general we can consider a single component of interest, where the dipoles of the emitters are aligned with the mode polarization.
Thus,  if $x_a>x_b$, then
$  G_W({\bf r}_a,{\bf r}_b,\omega)  = iAe^{i\omega \tau_{ab}}$ (where we choose the relative polarization component),
and $\tau_{ab}$ 
is the delay time to propagate from point ${\bf r}_a$ to point ${\bf r}_b$. 
Note that the wave vector is dispersive in general, so $k=k(\omega)$.

Using a photonic Green function approach for 
the waveguide system, 
a Lindblad master equation 
can be derived within 
a Markov approximation 
\cite{PhysRevA.91.051803,PhysRevA.66.063810}:
\begin{equation}
\begin{split}
    &\Dot{\rho}(t) = - \frac{i}{\hbar} \left[ H_{\rm S}^{\rm },\rho (t) \right] - i \sum_{n,n'}^{n \neq n'} \delta_{nn'} \left[ \sigma^+_n \sigma^-_{n'},\rho(t) \right] \\   
    &+ \sum_{n,n'} \frac{\Gamma_{nn'}}{2} \left[ 2 \sigma^+_{n'} \rho (t) \sigma^-_n - \rho (t) \sigma^-_n \sigma^+_{n'} - \sigma^-_n \sigma^+_{n'} \rho (t) \right],
\end{split}
\label{L_ME}
\end{equation}
where the various coupling rates will be defined below.
Notably, the waveguide is treated as a reservoir here, so waveguide photons are assumed to be uncorrelated with the
qubits,
and $\sigma_n^\pm$ are the usual Pauli operators for the 
qubits treated as TLSs,
each with a dipole moment
${\bf d}_n$.

To solve the Markovian master equation, we  first define the incoherent scattering rates  used in the Linbladian,
which are obtained from the medium Green function,
\begin{equation}
    \Gamma_{ab}|_{a\neq b} = \frac{2{\bf d}_a \cdot {\rm Im}{\bf G}({\bf r}_a,{\bf r}_b,\omega_b) \cdot {\bf d}_b }{\epsilon_0\hbar},
\end{equation}
\begin{equation}
\gamma_a \equiv \Gamma_{aa} = \frac{2{\bf d}_a \cdot {\rm Im}{\bf G}({\bf r}_a,{\bf r}_a,\omega_a) \cdot {\bf d}_a }{\epsilon_0\hbar},  %
\end{equation}
where the latter term is the usual spontaneous emission rate
from a single emitter, and the former term accounts
for inter-qubit photon transfer.
Using the waveguide Green functions, 
and considering three qubits (1-mirror, 2-probe,  and 3-mirror),  we have
\begin{equation}
\begin{split}
    &\Gamma_{12} = \Gamma_{21} = \sqrt{\gamma_m \gamma_p} \, {\rm Re} [e^{i \phi_{m_1,p}}] ,\\
    &\Gamma_{13} = \Gamma_{31} = \gamma_m {\rm Re}[e^{i \phi_{m_1,m_2}}],  \, \\
    &\Gamma_{23} = \Gamma_{32} = \sqrt{\gamma_m \gamma_p}  {\rm Re}[e^{i \phi_{m_2,p}}],
\end{split}   
\end{equation}
where $\phi_{m_1,p}$, $\phi_{m_2,p}$ and $\phi_{m_1,m_2}$ represent the phase between each mirror, the probe dot, and between both mirrors, respectively, due to their position in the waveguide. 
Explicitly,
$\phi_{n,n’} = k (x_n - x_{n'})$,
which is also frequency dependent
in general. 

The {\it coherent} coupling rates are obtained from the real part of the Green functions,
\begin{equation}
    \delta_{ab}|_{a\neq b} = - \frac{2{\bf d}_a \cdot {\rm Re}{\bf G}({\bf r}_a,{\bf r}_b,\omega_b) \cdot {\bf d}_b }{2\epsilon_0\hbar},
\end{equation}
which, for the waveguide system, simplifies  to
\begin{equation}
\begin{split}
    &\delta_{12} = \delta_{21} = \sqrt{\gamma_m \gamma_p}  \,{\rm Im}[e^{i \phi_{m_1,p}}] , \\
    &\delta_{13} = \delta_{31} = \gamma_m\,  {\rm Im}[e^{i \phi_{m_1,m_2}}] ,  \\
    &\delta_{23} = g_{32} = \sqrt{\gamma_m \gamma_p}  \,{\rm Im}[e^{i \phi_{m_2,p}}] .
\end{split}   
\end{equation}

The effective system Hamiltonian, including dipole-dipole coupling  mediated by the waveguide, is then
\begin{equation}
\begin{split}
    H_{\rm S}^{\rm eff} &= \frac{\sqrt{\gamma_m \gamma_p}}{2}\, {\rm Im}[e^{i \phi_{m_1,p}}] \left[ \sigma^+_{m_1} \sigma^-_{p} + \sigma^+_{p} \sigma^-_{m_1}  \right] \\
    &+  \frac{\sqrt{\gamma_m \gamma_p}}{2}\, {\rm Im}[e^{i \phi_{m_2,p}}]  \big[ \sigma^+_{m_2} \sigma^-_{p} + \sigma^+_{p} \sigma^-_{m_2}  \big] \\
    &+ \frac{\gamma_m}{2}\, {\rm Im}[e^{i \phi_{m_1,m_2}}] \big[ \sigma^+_{m_1} \sigma^-_{m_2} 
    + \sigma^+_{m_2} \sigma^-_{m_1}  \big] \\
    &+ \Omega \big[ \sigma^-_{p} + \sigma^+_{p} \big] + \Delta \big[ \sigma^+_{m_1} \sigma^-_{m_1} + \sigma^+_{p} \sigma^-_{p}  +\sigma^+_{m_2} \sigma^-_{m_2}  \big],
\end{split}
\label{hb}
\end{equation}
with a possible pump Rabi field $\Omega$ exciting  the probe qubit, and we also introduced a laser-qubit detuning, $\Delta=\omega_0-\omega_{L}$, 
where $\omega_0=\omega_a=\omega_b$,
and $\omega_L$ is the frequency of the pump. 

Thus, we can write the Markovian master equation as
\begin{equation}
\begin{split}
    &\Dot{\rho}(t) = - \frac{i}{\hbar} \left[ H_{\rm S}^{\rm eff},\rho (t) \right]  \\   
    &+ \sum_{n,n'} \frac{\Gamma_{nn'}}{2} \left[ 2 \sigma^+_{n'} \rho (t) \sigma^-_n - \rho (t) \sigma^-_n \sigma^+_{n'} - \sigma^-_n \sigma^+_{n'} \rho (t) \right].
\end{split}
\label{L_ME2}
\end{equation}

The  detected spectrum at $\mathbf{r}_D$ can be obtained 
from the the first-order quantum correlation function $G^{(1)}(\mathbf{r}_D,\tau)=\langle \hat{\mathbf{E}}^-(\mathbf{r}_D, t)\hat{\mathbf{E}}^+(\mathbf{r}_D, t+\tau)\rangle$.  In the rotating frame at the laser frequency, the total spectrum is
\begin{align}
&S^T_{D}(\omega)= \nonumber \\
& \lim_{T \to \infty}\frac{1}{T}\int_0^T dt\int_0^T dt' 
 \langle\hat{\mathbf{E}}^- (\mathbf{r}_D, t) \hat{\mathbf{E}}^+(\mathbf{r}_D, t')\rangle e^{ i(\omega_L - \omega ) (t-t') }.
\end{align}
 Inserting the formal solution for the electric field operator, obtained from Heisenberg's equation of motion\cite{PhysRevA.91.051803}, then 
\begin{equation}
\langle\hat{\mathbf{E}}^- (\mathbf{r}_D, \omega) \hat{\mathbf{E}}^+(\mathbf{r}_D, \omega)\rangle=\sum_{n, n'}g_{n, n'}(\omega)\langle{\sigma}_n^+(\omega) {\sigma}_{n'}^-(\omega)\rangle,
\label{eq:epem}
\end{equation}
where the emitter coupling term is
\begin{equation}
g_{n,n'}(\omega)=\frac{1}{\epsilon_0^2}\mathbf{d}_n\cdot\mathbf{G}^*(\mathbf{r}_{n}, \mathbf{r}_D, \omega)\cdot\mathbf{G}(\mathbf{r}_D, \mathbf{r}_{n'}, \omega)\cdot\mathbf{d}_{n'},
\end{equation}
and $g_{n',n}=g^*_{n, n'}$.

The incoherent spectrum can be separated into direct contributions and interference terms, so that
(${\bf r}_D$ is implicit):
\begin{align}
S_{D}(\omega)&=\sum_n\left |\mathbf{G}(\mathbf{r}_D, \mathbf{r}_n, \omega)\cdot\frac{\mathbf{d}_n}{\epsilon_0}\right|^2
 {\rm Re}\{ S^0_{ n, n}(\omega)\} \nonumber \\
& +\sum_{n, n'}^{n\neq n'}\text{Re}\{g_{n, n'}(\omega)S^0_{n, n'}(\omega)\},
\label{eq:Stot}
\end{align}
with 
\begin{align}
 S^0_{n, n'}(\omega)&=\lim_{t \to \infty} \int_0^\infty dt' ( \langle{\sigma}^+_n (t+t') {\sigma}^-_{n'}(t)\rangle \nonumber \\
 &-\langle{ \sigma }^+_n(t)\rangle\langle{\sigma}^-_{n'}(t)\rangle ) e^{ -i(\omega-\omega_L ) t' },
\end{align}
where the latter contribution subtracts the coherent spectrum, which is simply a Dirac delta function for continuous wave (CW) pumping.
A full derivation of the master equation and spectra are presented in Refs.~\onlinecite{PhysRevA.91.051803,PhysRevA.66.063810}.

We recognize that the total spectrum contains terms corresponding to the spectrum emitted from the single qubits ($n=n'$), as well as interference terms ($n\neq n'$).
Finally, to be consistent with the Markov approximation used in the master equations, we replace 
${\bf G}({\bf r}_D,{\bf r}_n,\omega)$ by ${\bf G}({\bf r}_D,{\bf r}_n,\omega_L)$, though this is not a model requirement.

\subsection{Matrix products states}
\label{subsection:MPS}

For our MPS approach, we can write our state as follows,
\begin{equation}
\label{waveguidecase}
    \ket{\psi}=\sum_{i_s i_1...i_N} A_{a_1}^{i_s}A_{a_1,a_2}^{i_1} \hdots A_{a_{N-1},a_{N}}^{i_{N-1}}A_{a_{N}}^{i_{N}}\ket{i_s, i_1,\hdots,i_N},
\end{equation}
where $i_s$ represents the system bin containing the 3 TLSs, and the remaining $i_1,\hdots,i_N$ terms represent the discretized waveguide. Here, each of the $A$ terms is a tensor where the subscripts $a_1,\hdots,a_{N-1}$ are the auxiliary dimensions of each element and the superscripts $i_1,\hdots,i_N$ represent the physical dimensions of the system. Further information about the MPS method can be found in \cite{cajitas}.

Setting the units such that $\hbar=1$ for convenience, we
consider the total Hamiltonian, 
\begin{equation}
    H = H_{\rm sys} + H_{\rm B} + H_{\rm int},
\end{equation}
where 
\begin{equation}
\begin{split}
    H_{\rm sys} = \sum_{n=m_1,p,m_2} \Big[ \omega_n \sigma^+_n \sigma^-_n
    - \frac{1}{2} \left( \Omega_n \sigma^-_n e^{i\omega_L t} + {\rm H.c.} \right) \Big],
    \label{Hsys}
\end{split}
\end{equation}
\begin{equation}
    H_{\rm B} = \sum_{i = L,R} \int_B d\omega \hbar \omega b_i^\dagger (\omega) b_i(\omega),
\end{equation}
\begin{equation}
\begin{split}
    H_{\rm int} = i \sum_{i,n} \int_B d\omega \Big[\kappa_i (\omega) b_i^\dagger(\omega)\sigma^-_n e^{-i\omega x_i/v_i} 
    - {\rm H.c.} \Big]\,.
\end{split}
\end{equation}

 Switching to the interaction picture with respect to the bath Hamiltonian and into a rotating frame with the frequency $\omega_L$, 
then 
\begin{equation}
    H_{\rm sys} = \sum_{n=m_1,p,m_2} \left[ - \Delta_n \sigma^+_n \sigma^-_n - \frac{1}{2} \left( \Omega_n \sigma^-_n +
    {\rm H.c.} \right) \right],
\end{equation}
with $\Delta_n = \omega_L - \omega_n$ 
Next, choosing $\kappa_i \to \sqrt{\gamma_i / 2\pi}$, then 
\begin{equation}
\begin{split}
    H_{\rm int} = &\frac{i}{\sqrt{2\pi}} \sum_{i,n} \int_B d\omega \\
    & \Big[\sqrt{\gamma_i} b_i^\dagger(\omega) \sigma^-_n e^{-i\omega x_i/v_i} e^{i(\omega - \omega_L)t} 
    - {\rm H.c.} \Big]\,.
\end{split}    
\end{equation}
If we choose $x=0$ for the middle dot, then we can choose $x_1 = -x$, $x_2=0$ and $x_3=x$. The group velocity of the waveguide mode is considered constant (over the bandwidth of interest), with $v_L = -v$ and $v_R = v$.

The delay time between the probe dot and the mirror dot ($\tau_{pm}$ for our symmetric system) is redefined as $\tau_{pm} \equiv \tau_m $ for a simpler notation, and $\tau_m = x/v$. The boson operators in the time domain are
\begin{equation}
\begin{split}
    &b_i^\dagger(t) = \frac{1}{\sqrt{2\pi}} \int d\omega b_i^\dagger(\omega) e^{i(\omega - \omega_L)t}, \\
    &b_i^\dagger(t-\tau_m) = \frac{1}{\sqrt{2\pi}} \int d\omega b_i^\dagger(\omega) e^{i(\omega - \omega_L)(t-\tau_m)}, \\
    &b_i^\dagger(t+\tau_m) = \frac{1}{\sqrt{2\pi}} \int d\omega b_i^\dagger(\omega) e^{i(\omega - \omega_L)(t +\tau_m)}, 
\end{split}    
\end{equation}
and corresponding equations for $b_i(t)$, $b_i(t-\tau_m)$ and $b_i(t+\tau_m)$.
Thus, we can write
\begin{equation}
\begin{split}
    &-i{H_{\rm int}} = \\
    &\sqrt{\gamma_L} b_L^\dagger(t - \tau_m) \sigma^-_{m_1} e^{i\omega_L \tau_m} 
    + \sqrt{\gamma_R} b_R^\dagger(t + \tau_m) \sigma^-_{m_1} e^{-i\omega_L \tau_m} \\
    &- \sqrt{\gamma_L} b_L(t - \tau_m) \sigma^+_{m_1} e^{-i\omega_L \tau_m}
    + \sqrt{\gamma_R} b_R(t + \tau_m) \sigma^+_{m_1} e^{i\omega_L \tau_m}\\
    &+ \sqrt{\gamma_L} b_L^\dagger(t) \sigma^-_p + \sqrt{\gamma_R} b_R^\dagger(t) \sigma^-_p\\
    &- \sqrt{\gamma_L} b_L(t) \sigma^+_p + \sqrt{\gamma_R} b_R(t) \sigma^+_p\\
    &+ \sqrt{\gamma_L} b_L^\dagger(t + \tau_m) \sigma^-_{m_2} e^{-i\omega_L \tau_m}
    + \sqrt{\gamma_R} b_R^\dagger(t - \tau_m) \sigma^-_{m_2} e^{i\omega_L \tau_m} \\
    &- \sqrt{\gamma_L} b_L(t + \tau_m) \sigma^+_{m_2} e^{i\omega_L \tau_m}
    + \sqrt{\gamma_R} b_R(t - \tau_m) \sigma^+_{m_2} e^{-i\omega_L \tau_m}.
\end{split}    
\label{H_int2}
\end{equation}

Finally, we can redefine the boson operators, choosing $t + \tau_m \to t'$, and we define the phases between dots as in Sect.~\ref{subsection:ME} (i.e., $\omega_L \tau_m = \phi_{m1,p} = \phi_{m1,p} = \phi_{m,p}$ phase between a mirror qubit and the probe qubit, and $\omega_L \tau_m = \phi_{m1,m2}$ phases between mirrors).  
Using these in Eq.~\eqref{H_int2}, we get to the final equation for the interaction Hamiltonian (and we drop the prime superscript in $t'$),
\begin{equation}
\begin{split}
    &H_{\rm int}/i = \sqrt{\gamma_L} b_L^\dagger(t - 2\tau_m) \sigma^-_{m_1} 
    + \sqrt{\gamma_R} b_R^\dagger(t) \sigma^-_{m_1} e^{-i\phi_{m1,m2}} \\
    &- \sqrt{\gamma_L} b_L(t - 2\tau_m) \sigma^+_{m_1} 
    + \sqrt{\gamma_R} b_R(t) \sigma^+_{m_1} e^{i\phi_{m1,m2}} \\
    &+ \sqrt{\gamma_L} b_L^\dagger(t-\tau_m) \sigma^-_p e^{-i\phi_{m,p}}+ \sqrt{\gamma_R} b_R^\dagger(t-\tau_m) \sigma^-_p e^{-i\phi_{m,p}}\\
    &- \sqrt{\gamma_L} b_L(t-\tau_m) \sigma^+_p e^{i\phi_{m,p}} + \sqrt{\gamma_R} b_R(t-\tau_m) \sigma^+_p e^{i\phi_{m,p}}\\
    &+ \sqrt{\gamma_L} b_L^\dagger(t) \sigma^-_{m_2} e^{-i\phi_{m1,m2}} 
    + \sqrt{\gamma_R} b_R^\dagger(t - 2\tau_m) \sigma^-_{m_2}  \\
    &- \sqrt{\gamma_L} b_L(t) \sigma^+_{m_2} e^{i\phi_{m1,m2}}
    + \sqrt{\gamma_R} b_R(t - 2\tau_m) \sigma^+_{m_2}.
\end{split}    
\label{H_int3}
\end{equation}

The spectrum of the output field, just after the right mirror dot, in steady state, is obtained from  
\begin{equation}
    S(\omega) \to 2 \mathcal{R} \int_0^\infty dt' \langle b^{\dagger}(t) b(t-t') \rangle e^{i(\omega-\omega_L) t'},
\end{equation}
 which can be written in the discrete time bin scheme \cite{cajitas}, as
\begin{equation}
    S_{i,j}(\omega) \to 2 \mathcal{R} \frac{1}{\Delta t} \sum_{p=0}^{M-1} \langle \Delta B^{\dagger}(t_q) \Delta B(t_{q-p}) \rangle e^{i(\omega-\omega_L) p \Delta t},
\end{equation}
where $q=k_{\rm max} - l - 1$ ($l = \tau / \Delta t$ and $k_{\rm max}$ last time bin), and $p \in \{0,M-1 \}$ with $M$ large enough to resolve the spectrum. This value will vary for each specific simulation case, depending on the required $\Delta t$, and the necessary time to resolve the photon correlation. In general, we will need a smaller $\Delta t$ for a small retardation and a large pump, and a longer correlation time ($t_{\rm cor}$) for a large retardation. Note that $M = 1600$ is a typical value used in our calculations for an intermediate case where $\Delta t = 0.025 \gamma_p$ and $t_{\rm cor} = 40 \gamma_p$.

\subsection{Exact solution for the probe-qubit electric field under linear response}

For reference, we briefly discuss the linear regime solution~\cite{PhysRevA.104.L031701}.
 Defining the distance between the mirror dots as
$L$,  it is possible to derive an exact solution
for the scattered electric field at the probe qubit in a weak excitation approximation (namely, under linear response)
\begin{align}
    \tilde{E}_{\rm s}({\bf r}_p,\omega) &= 
     \frac{i \omega_p\tilde\gamma_p}{\omega_p^2-\omega^2-i\omega \tilde\gamma_p},
     \label{eq:E3_exact}
    \end{align}
where all qubits are on-resonance and the modified decay rate is defined from~\cite{PhysRevA.104.L031701}
\begin{align}
\tilde\gamma_p =
\gamma_p &\bigg[1+  e^{ikL}r_1(\omega)+
(e^{ikL/2} +  e^{ikL/2} e^{ikL}r_1(\omega)
) \times \nonumber \\
& \ \ \frac{r_1(\omega)\left(e^{ikL/2}+r_1(\omega)e^{ikL/2}e^{ikL}\right)} {1-r_1^2(\omega)e^{2ikL} } \bigg], 
\label{eq:gamma3_exact}
\end{align}
and we use the single qubit reflection coefficient
\begin{equation}
r_1(\omega) = \frac{{\bf E}_{\rm r}(\mathbf{r}; x\rightarrow -\infty)}{{{\bf E}^{\rm h}}(\mathbf{r}; x\rightarrow -\infty)}
=  \frac{i\omega_0\gamma e^{i\phi(x_d)}}
{\omega_0^2-\omega^2-i\omega_0
\gamma},
\label{r:1dot}
\end{equation}
where $\phi(x_d)$ is  a positional dependent phase. 
Using the mirror qubits,
then $\gamma=\gamma_m$.
Clearly this solution contains multiple resonances. With an appropriate choice for the decay rates, one can achieve an analogue of vacuum Rabi splitting for the main two polariton peaks, when $g>\gamma_m$,
where 
$g=\sqrt{2\gamma_m\gamma_p}/2$. 
However, this limit is only achieved in the Markovian regime. In the same limit,
the broadening of these peaks is 
$\gamma_p/4$, which can be compared
with $\kappa/2$ from the usual cavity-QED system in the absence of vertical (background) decay~\cite{PhysRevB.60.13276}.

\section{Results}
\label{sec:results}

\subsection{Linear response of the vacuum Rabi polariton poles for different delay times}

\begin{figure}[ht]
    \centering
    \includegraphics[trim=0cm 0cm 0cm 0cm, clip,width=0.9 \columnwidth]{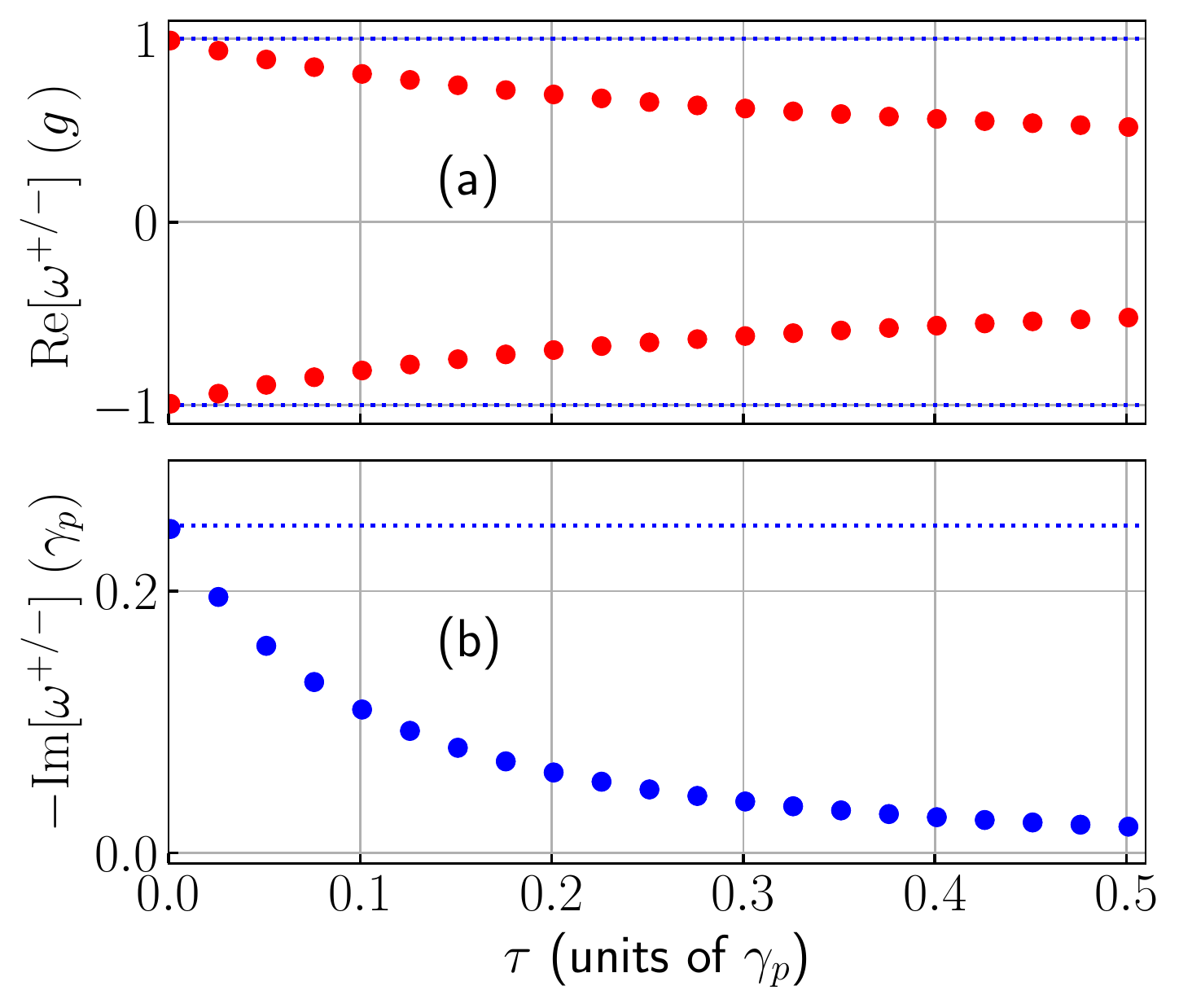}
    \caption{
    Complex poles of the two main polariton states calculated using Eq.~\eqref{eq:E3_exact}, with $\tau=2\tau_{m}$ the delay time between mirror qubits, and $\gamma_m=10\gamma_p$. The dashed lines show the non-retarded solution (Markov limit). 
    } 
    \label{fig:roots}
\end{figure}

As  shown in Ref.~\cite{PhysRevA.104.L031701}, retardation effects can have a significant influence on the coupling rate of this cavitylike system. 
This linear result is shown  in Fig.~\ref{fig:roots}, where the first near-resonant complex poles are calculated as a function of the delay time. 
However, note that the general solution contains multiple resonances.
We observe how the values of the complex poles decrease for an increasing retardation time.
The frequency units are shown in terms of an effective cavity-atom coupling rate, $g$. We stress that this coupling rate is only valid for small delay times.
In general, the resonances depend on the distance between the qubits. Hence, later will  consider an effective coupling rate $g_{\rm eff} < g$, when one enters a non-Markovian regime, as the Rabi doublets decrease in energy, and spectrally sharpen. 

It is important to realize that these non-Markovian delay times correspond to much longer length scales than a few wavelengths.
For example, 
if we consider a delay time of $\tau =0.2/\gamma_p $, with a typical QD decay rate $\gamma_p=1 \rm\, ns^{-1}$
and a group index $n_g=c/v_g=10$, the distance between the probe qubit and the mirror one will be
6 microns, which is thousands of wavelengths for typical 
integrated QDs (e.g.,
with a wavelength of 1000 nm). Thus, low loss waveguides are required, such as SiN~\cite{Mnaymneh2019}.

\subsection{Nonlinear Markovian regime}
\subsubsection{Emitted Spectrum for Different Qubit Decay Rate Ratios}
\begin{figure} [ht]
\centering
\includegraphics[width=\columnwidth]{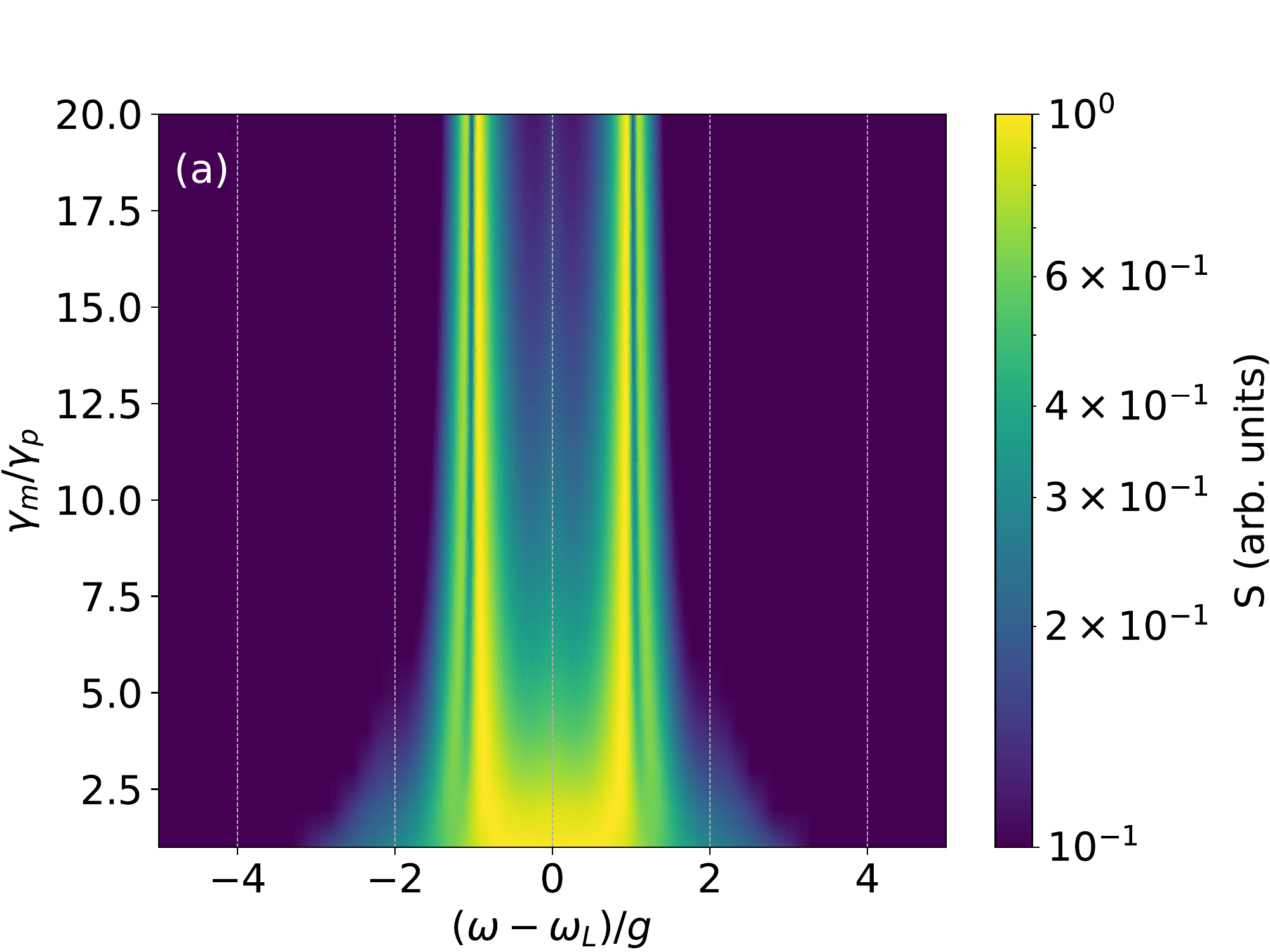} \\
\includegraphics[width=\columnwidth]{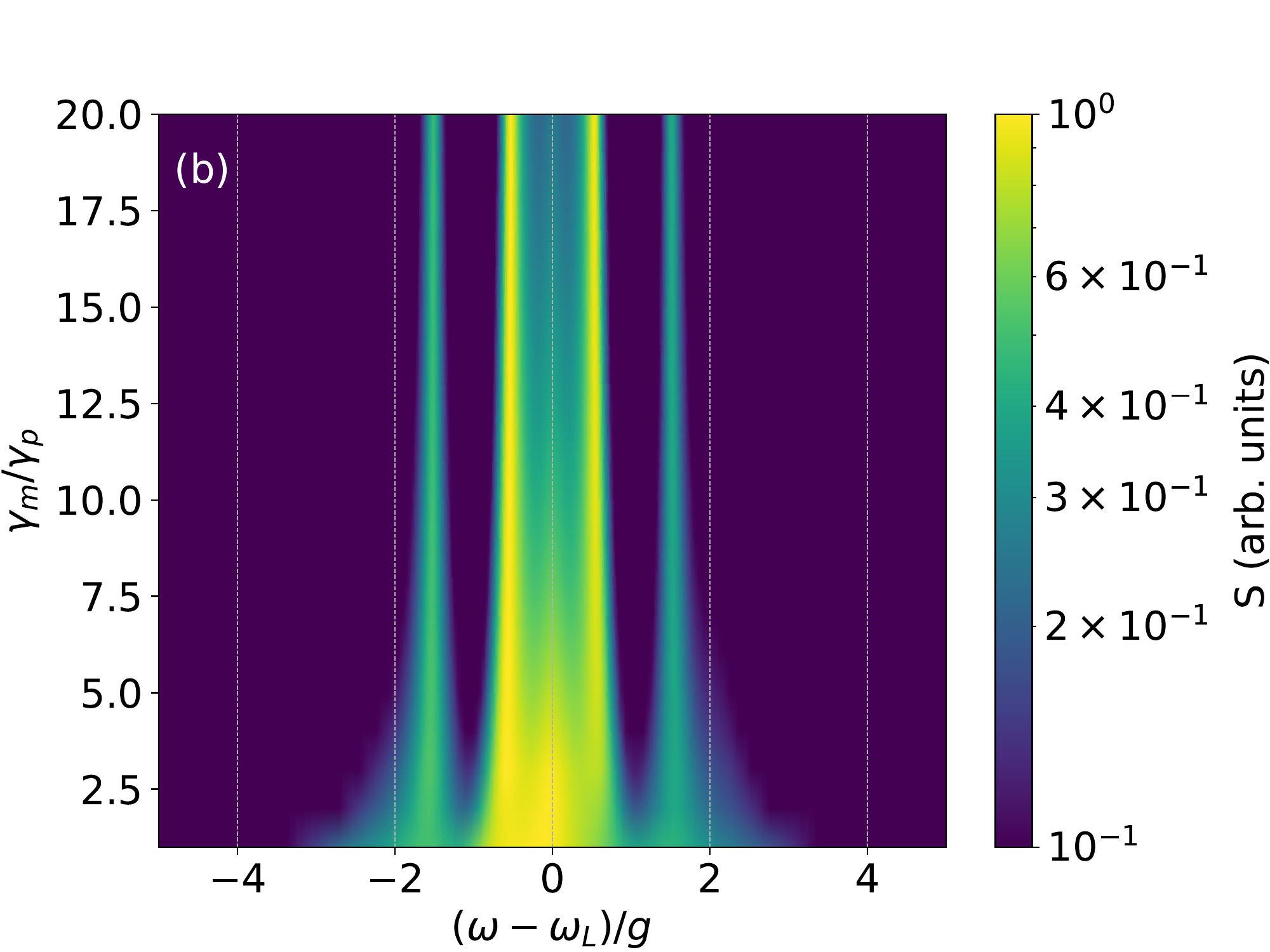}%
 \caption{Master equation calculation of the total spectrum of a three-qubit system for different ratios of $\gamma_m / \gamma_p$ in the Markovian (non-retarded) regime. Probe dot pumped with $\Omega=0.5 \gamma_p \approx 0.22 g$. (a) On-resonance pumping. (b) Off-resonance with $\Delta_n = g/2$.%
 }
 \label{gamma_rate}
\end{figure}
\begin{figure*}[t]
    \centering
    \includegraphics[trim={1cm 2cm 1cm 1cm},clip,width=0.9\textwidth]{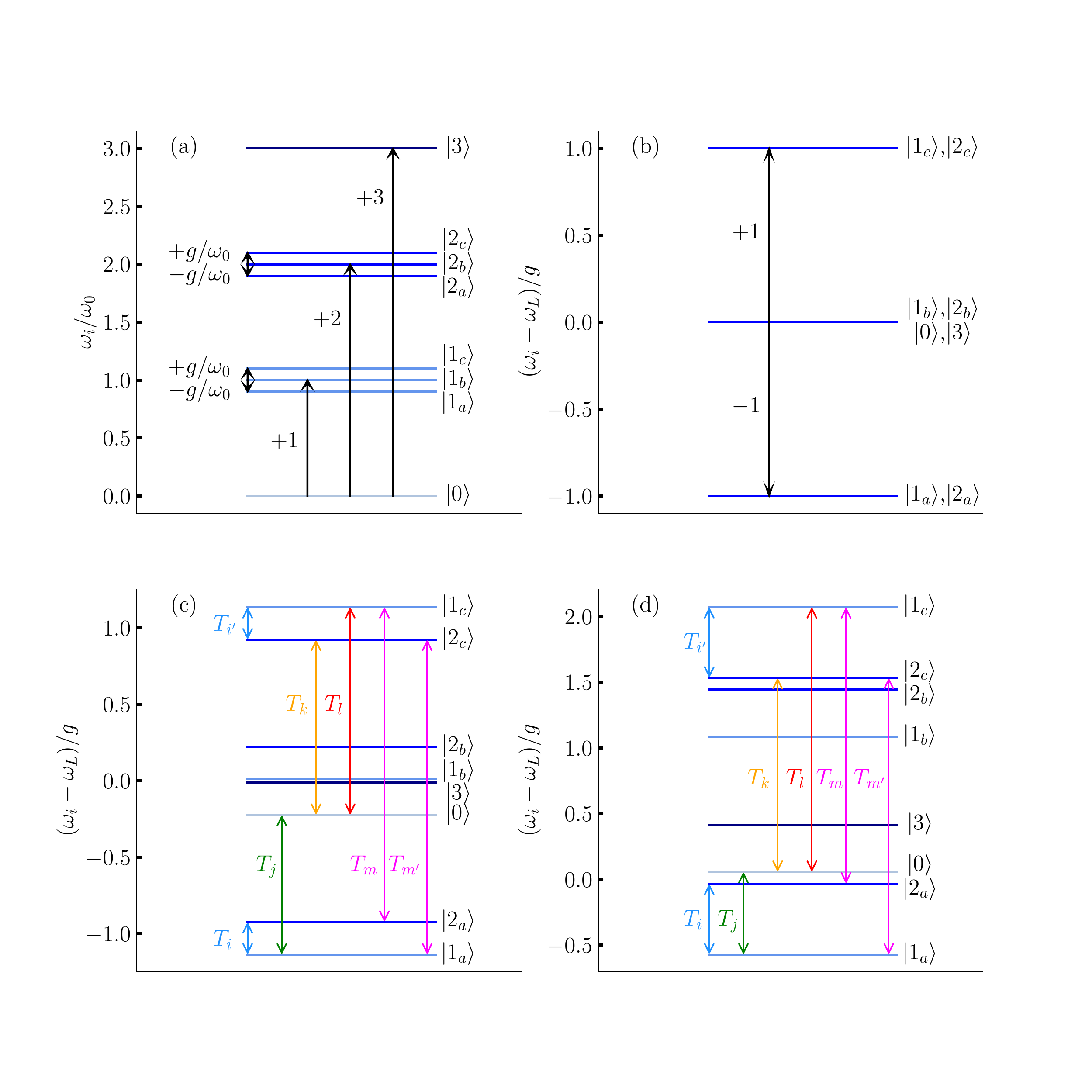}
    \caption{Energy levels of the three-qubit system in a Markovian regime: (a) Dressed-state basis in the absence of an optical pumping without using the interaction picture. (b) Same case as (a) but now in the interaction picture. (c) Optically driven case with  $\Omega = 0.5 \gamma_p \approx 0.22 g$ (on-resonance)  (d) Optically driven case with  $\Omega = 0.5 \gamma_p \approx 0.22 g$ and $\Delta = g_0/2$ (off-resonance). Prime labelled transitions are degenerate with their corresponding unprimed transition. This schematic only applies to the Markovian system, and in MPS, there are also multi-photon states.
    }
    \label{fig:energylev}
\end{figure*}
We next solve the three-qubit Markovian master equation [Eq.~\eqref{L_ME}] and investigate the analogue of a cavity-QED system by using the side qubits acting as dipole  mirrors \cite{Mirhosseini2019} (through resonant scattering), but now with optical pumping. All the calculations presented are performed in {\sc Python}, and for this first method, we make use of the QuTiP library \cite{johansson2012qutip,johansson_qutip_2013}. For the excitation, we pump the center TLS (probe qubit) with a sufficiently strong field, $\Omega=0.5\gamma_p$, so as to induce nonlinear interactions, and study the influence of the ratio $\gamma_m / \gamma_p$. 
Using resonant pumping, in Fig.~\ref{gamma_rate}(a), we show how, by increasing the $\gamma_m / \gamma_p$ ratio, the total spectrum [Eq.~\eqref{eq:Stot}] of our system resembles the characteristic Rabi splitting from a cavity system when pumped on-resonance (as also shown in Fig.~\ref{fig:roots}), under linear response. Note that if we go to the opposite limit, where $\gamma_m / \gamma_p \approx 0$, then the system behaves as a single TLS, since the coupling of the side dots in this limit is negligible \cite{doi:10.1126/science.1181918}. 

Based on these solutions, we choose a ratio of $\gamma_m / \gamma_p = 10$ for the rest of our investigations. This regime has also been experimentally demonstrated~\cite{Mirhosseini2019},
though our findings below are quite general.

Within the Markovian limit of the model, we see four resonances that show a splitting near the expected one-photon JC resonances (first photon ladder states, near $\pm g$). This indicates that we are  beyond the weak excitation limit where nonlinear effects appear from the pump field. 
The origin of these multiple resonances will be explained below in terms of the dressed states [Sect.~\ref{sec:dressed-subsubsection}].
 
Next, in Fig.~\ref{gamma_rate}(b), we show a similar study where the three qubits are still on-resonance with each other, but now we have an off-resonant pumping, with $\Delta_n=g/2$. 
In this case, we can again see four resonances, but they  now move farther from the expected linear one photon resonances due to the detuning introduced in the system.  We now clearly see additional nonlinear states that are not associated with the linear polariton states at $\pm g$.

\begin{center}
\begin{table}
\begin{tabular}{|c|c|} 
    \hline
    Dressed State &  Correspondence in bare basis \\
    \hline
    $\ket{0}$ &    $\ket{ggg}$ \\ 
    \hline
    $\ket{1_a} $ &  $\frac{1}{\sqrt{2}} \ket{geg} - \frac{1}{2} \left[ \ket{egg} + \ket{gge} \right]$  \\
    \hline
    $\ket{1_b}$  &  $\frac{1}{\sqrt{2}} \left[ \ket{egg} - \ket{gge} \right]$   \\ 
    \hline
    $\ket{1_c}$  &  $\frac{1}{\sqrt{2}} \ket{geg} + \frac{1}{2} \left[ \ket{egg} + \ket{gge} \right]$  \\ 
    \hline
    $\ket{2_a}$  &   $\frac{1}{\sqrt{2}} \ket{ege} - \frac{1}{2} \left[ \ket{eeg} + \ket{gee} \right]$  \\ 
    \hline
    $\ket{2_b}$  &   $\frac{1}{\sqrt{2}} \left[ \ket{gee} - \ket{eeg} \right]$ \\ 
    \hline
    $\ket{2_c}$  &  $\frac{1}{\sqrt{2}} \ket{ege} + \frac{1}{2} \left[ \ket{eeg} + \ket{gee} \right]$ \\ 
    \hline
    $\ket{3}$  &  $\ket{eee}$ \\ 
    \hline
\end{tabular}
\caption{\label{table1}Dressed state labels with coherent qubit interactions and the correspondence in terms of the bare states.}
\end{table}
\end{center}    

\subsubsection{Dressed State Picture}\label{sec:dressed-subsubsection}

To better explain the additional resonances that appear in the nonlinear regime, it is useful to use a dressed-state basis~\cite{PhysRevA.69.053810,Gustin:18}.
In the bare basis, we have the following states,
$\ket{ggg},$ $\ket{egg},$ $\ket{geg},$ $\ket{eeg},$ $\ket{gge},$ $\ket{ege},$ $\ket{gee},$ $\ket{eee}$, where `$g$'  and `$e$' correspond to the ground and excited levels, respectively, of the mirror-1 qubit, probe qubit and mirror-2 qubits. In the presence of coherent qubit interactions, we can obtain the dressed states from Eq.~\eqref{hb}, as shown and labelled in Table~\ref{table1}.
These are the natural dressed-states in the absence of any optical pumping, mediated by the coherent coupling between the qubits. Optical pumping will cause additional dressing. 

\begin{figure}[h]
    \centering
    \includegraphics[width=0.5\textwidth]{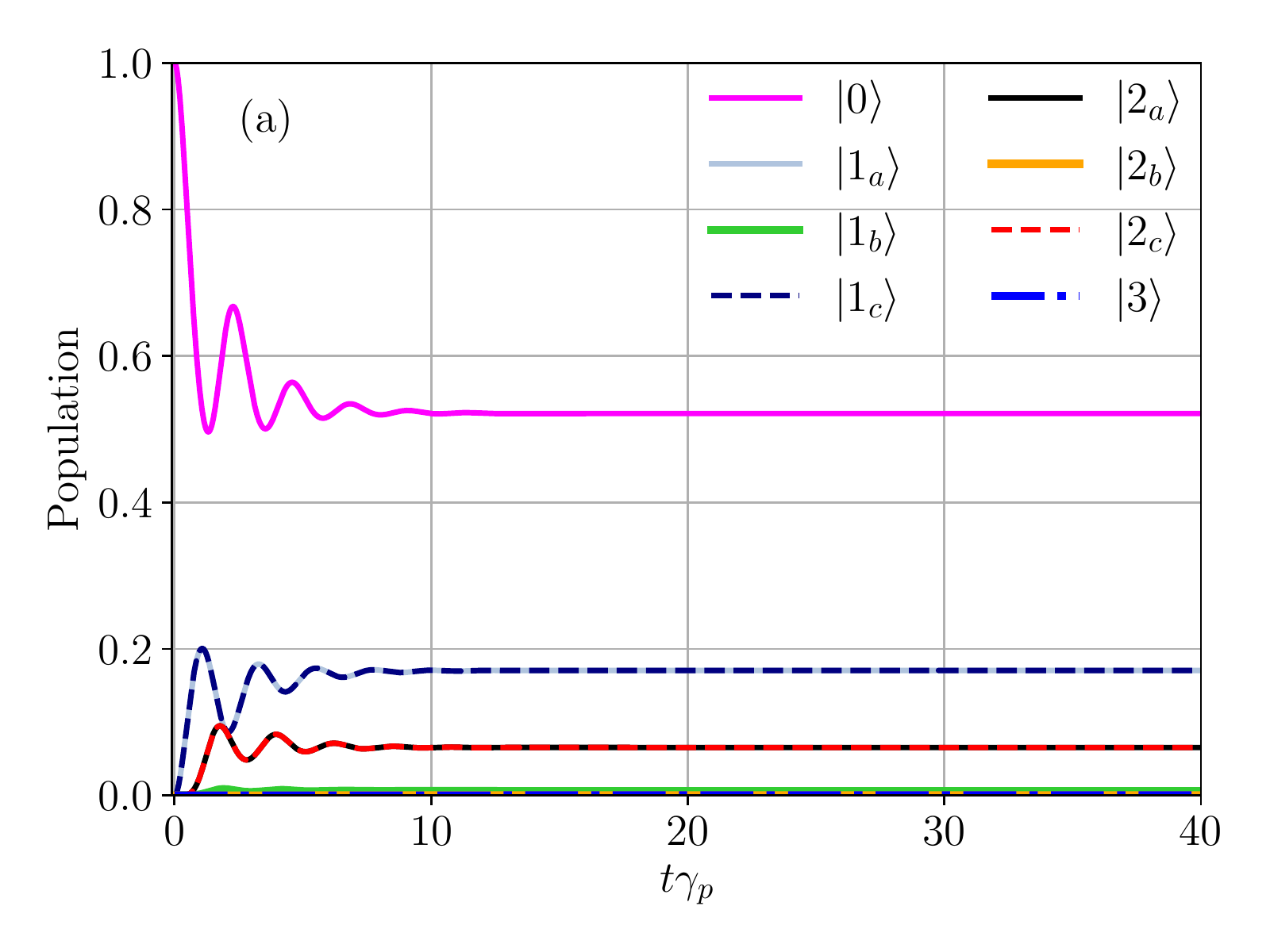}
    \includegraphics[width=0.5\textwidth]{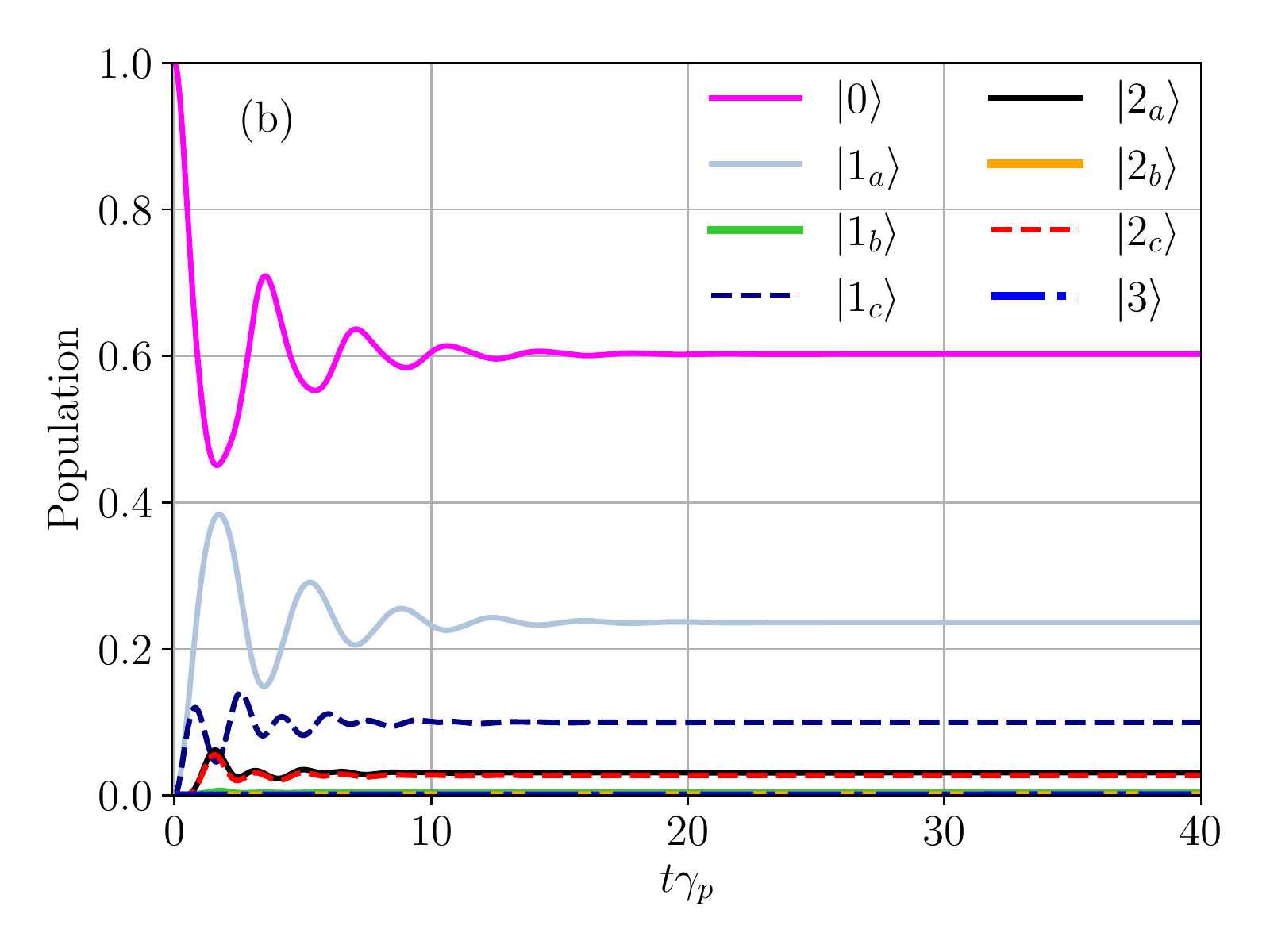}
    \caption{Dressed states population for $\Omega=0.5g_0$ (a) on resonance and (b) off resonance with a detuning of $\Delta_n = g/2$.}
    \label{fig:pop}
\end{figure}

In Figs.~\ref{fig:energylev}(a) and (b), we show the energy levels in the natural (unpumped) 
dressed state basis without using the interaction picture and its equivalence within the interaction picture. Then, in Figs.~\ref{fig:energylev}(c) and (d), we demonstrate how the states are additionally dressed by the optical pump field (on and off resonance, respectively) with their corresponding energy levels shifted. Key optical transitions are shown with arrows.

\begin{figure} [htb]
\centering
\includegraphics[width=\columnwidth]{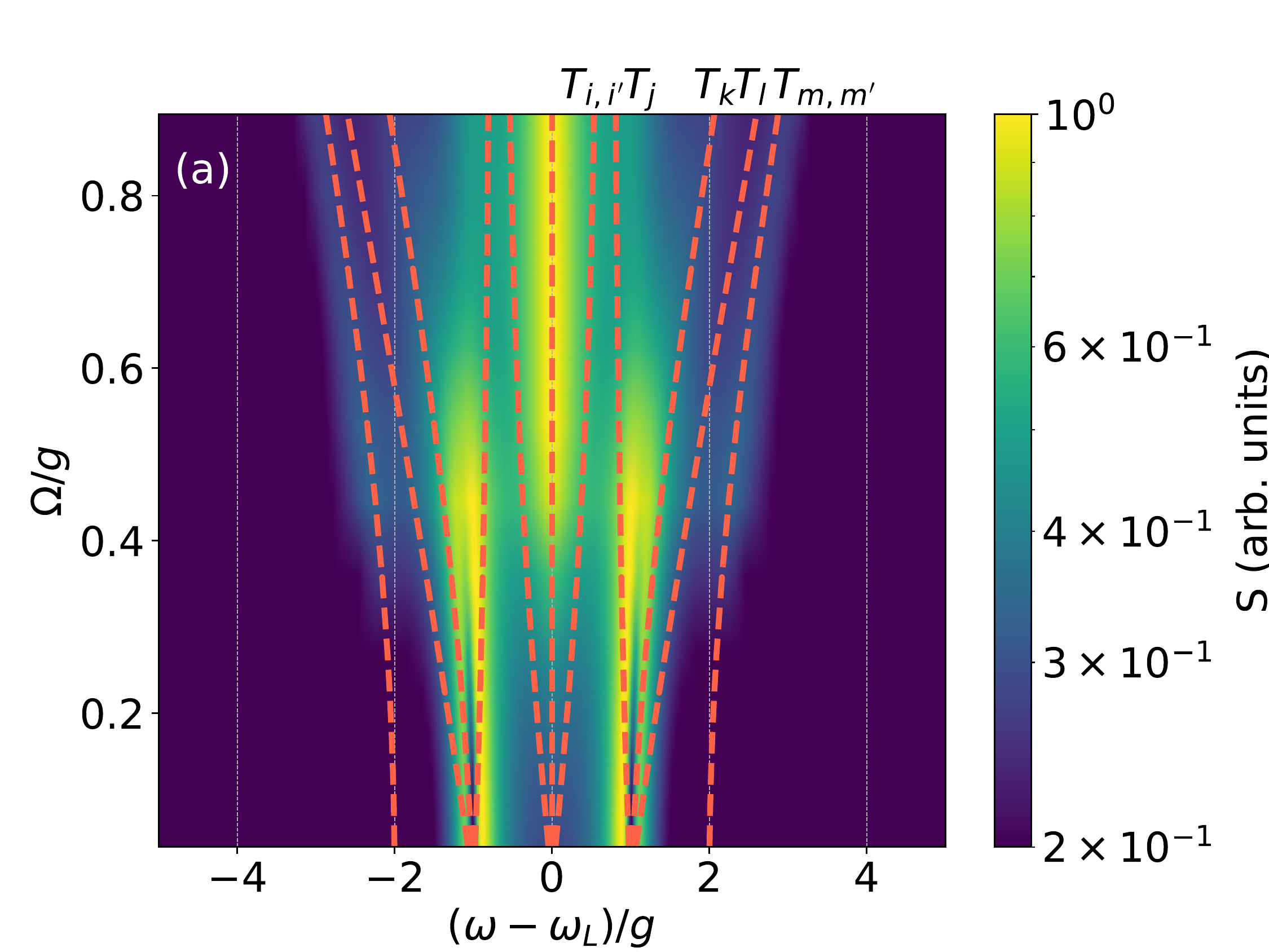} \\
\includegraphics[width=\columnwidth]{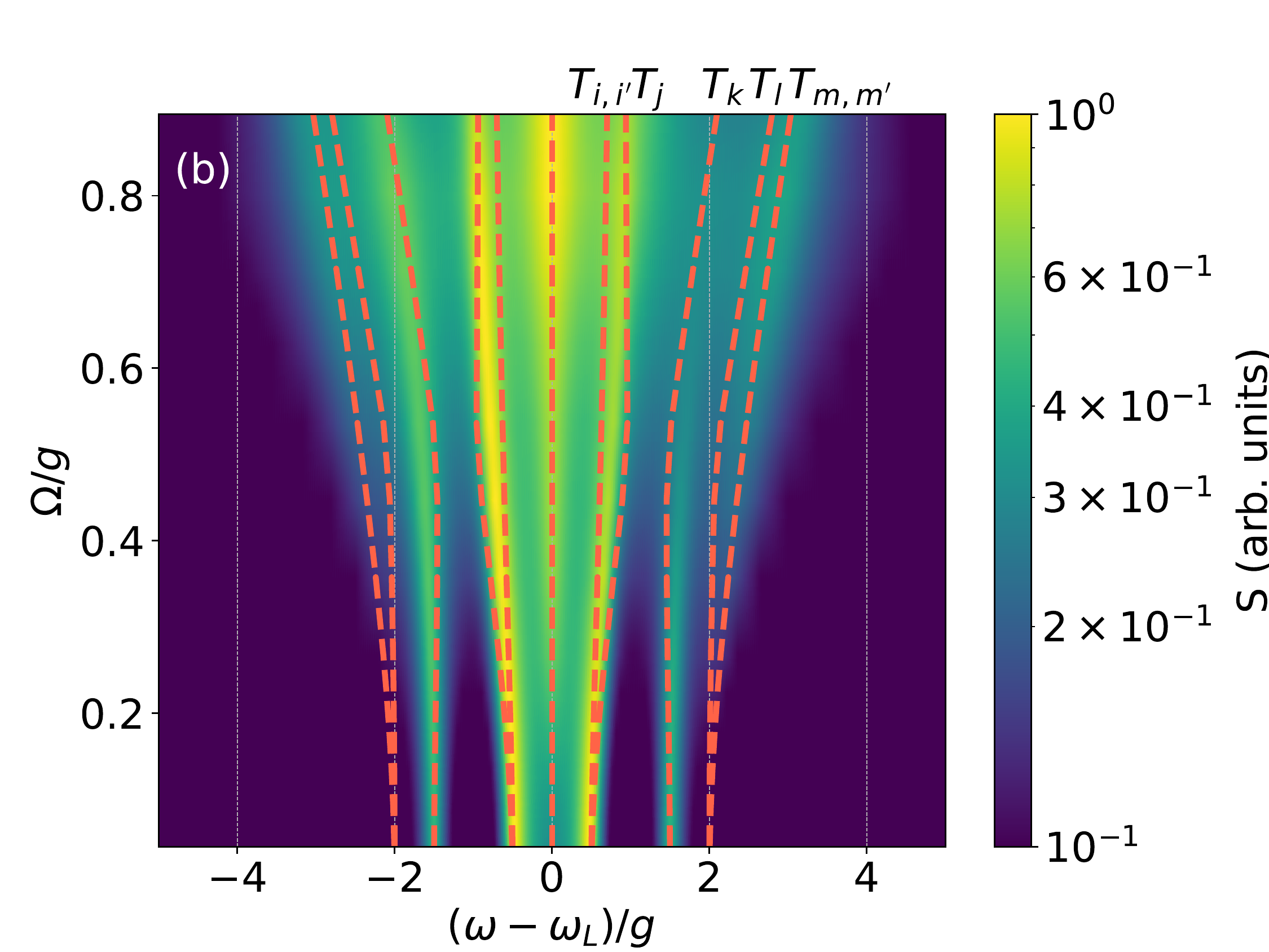}%
 \caption{Master equation calculations (Eq.~\ref{eq:Stot}) of the output spectrum of a three-qubit system for different pumping rates of the probe dot ($\gamma_m / \gamma_p = 10$). (a) On-resonance pumping. (b) Off-resonance with $\Delta_n = g/2$. Red dashed lines correspond to the transition lines shown in Fig.~\ref{fig:energylev}, and $T_{m, m'}$ and $T_{i,i'}$ correspond to degenerate transitions $T_{m} = T_{m'}$ and $T_{i} = T_{i'}$, respectively.
 }
 \label{qutip}
\end{figure}

We also examine the emitted spectra for different pump strengths and relate the spectral peaks 
with the possible energy transitions between these dressed states. In order to identify the possible energy transitions that are optically allowed (from a fairly high number), we first calculated the dressed-state populations in the natural dressed-state basis,
as shown in Fig.~\ref{fig:pop}. In both the resonant and detuned cases, we find five populated states, which allow us to identify the dominant transitions, which are then plotted with the numerically computed spectra shown in Fig.~\ref{qutip} (dashed curves). 

Figure~\ref{qutip}(a) shows the spectra for on-resonance excitation. If we focus on the positive (blue) frequency side of the spectra, we can distinguish five different peaks that are labelled as $T_i$, $T_{i^\prime}$, $T_j$, $T_k$, $T_l$, $T_m$ and $T_{m^\prime}$. These correspond directly to the transitions shown in Fig.~\ref{fig:energylev}(c) and (d), similarly labelled, plus the primed transitions which correspond to the degenerate states. Clearly, the correspondence between the full spectral results and the identified dressed-state resonances is very good.

Figure~\ref{qutip}(b) shows the case in which the laser is off-resonance with respect to the qubits, with a detuning of $\Delta_n = g/2$. A similar dressed energy-level ladder states scheme to Fig.~\ref{qutip}(a) is seen, but the position of the peaks observed in the spectrum are now qualitatively different due to different nonlinear dressing. However, the same basic photon transitions are identified in both scenarios. 

\subsubsection{Driven Jaynes-Cummings System}

It is useful to also compare the nonlinear response of the three qubit system with a driven JC system, since both these systems have a similar linear response. Moreover, a key aspect of using mirror qubits is that they are Fermionic systems, and one can expect significantly different nonlinear interactions, even when they are behaving as mirrors for cavity-QED.
Thus, below we show the main features of a (dissipative) JC model which solves a 
single TLS interacting with a single quantized electromagnetic field mode, in the usual rotating wave approximation  \cite{JaynesCummings}.
\begin{figure} [th]
\centering
\includegraphics[width=\columnwidth]{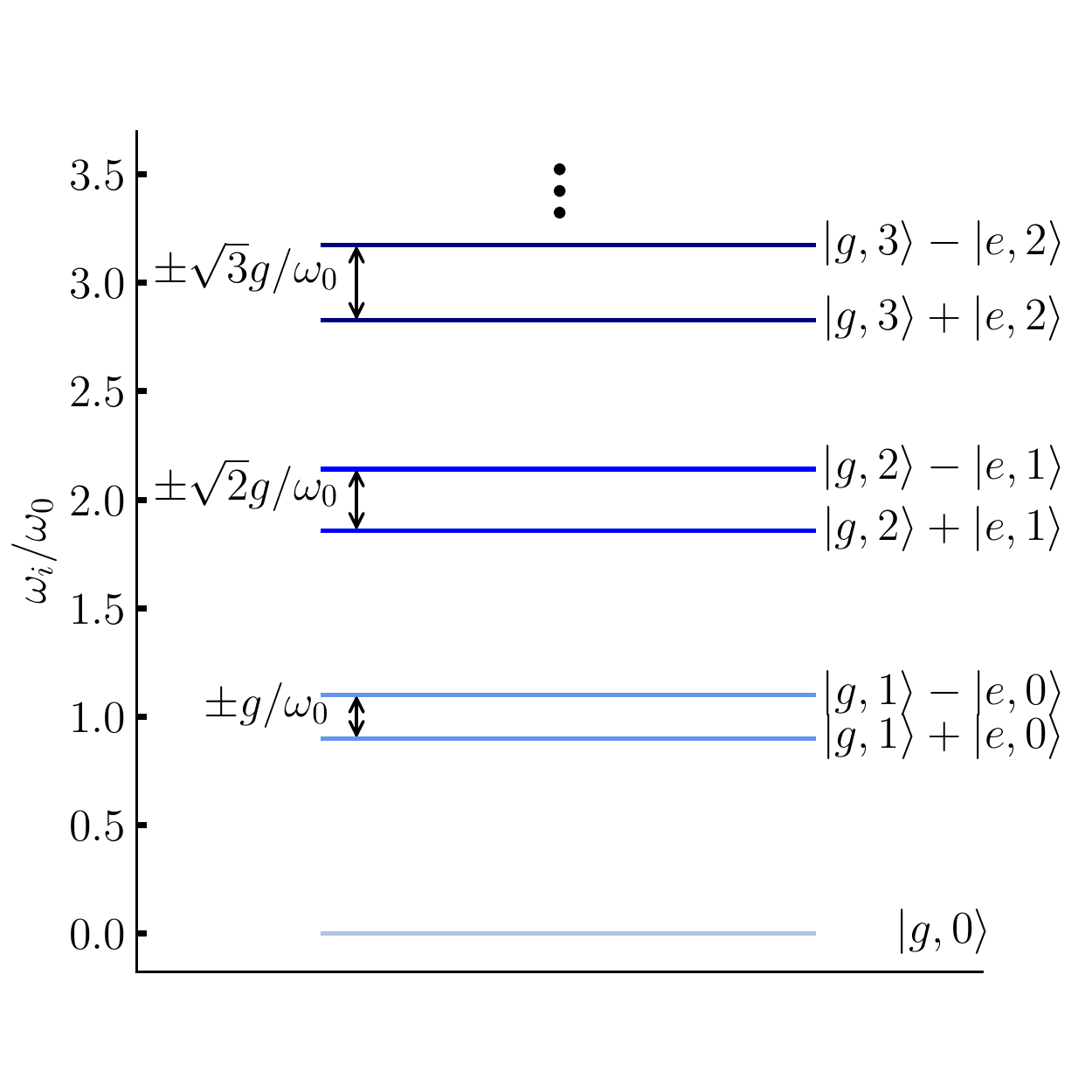}%
 \caption{Jaynes-Cummings model energy levels (of the first photon-matter states) where $\omega_i$ is the frequency of the $i^{\rm th}$ eigenenergy. The qubit and cavity are on resonance. The one photon states yield a similar strong coupling regime to the three qubit system, as shown in Fig.~\ref{fig:energylev}(a).
 }
 \label{JC_Model_en}
\end{figure}

For this model, we solve the corresponding master equation, following a similar approach to the one described in Sec.~\ref{subsection:ME}, where now the effective system Hamiltonian is
\cite{PhysRevA.40.5516}
\begin{equation}
\begin{split}
    H_{\rm S}^{\rm eff} &= 
    \Delta_a \big[ \sigma^+ \sigma^- \big] +
    \Delta_c \big[ a^{\dagger} a \big]  \\
    &+ g  \big[ a \sigma^+ + a^{\dagger} \sigma^- \big]
    + \Omega \big[ \sigma^- + \sigma^+ \big],
\end{split}
\label{hJC}
\end{equation}
with $\sigma^+$ and $\sigma^-$ the creation and annihilation operators of the single qubit, $a^{\dagger}$ and $a$ the (bosonic) ladder operators of the cavity, and $g$ the 
qubit-cavity coupling. As in the previous examples, $\Omega$ is the drive strength of the qubit and, in this JC case, $\Delta_a$ and $\Delta_c$ represent possible detuning between the drive and the atom or cavity, respectively. 
This is a standard driven JC model with an optical pumping term. In addition, we also include one collapse operator for the cavity mode decay,
\begin{equation}
    C=\sqrt{\kappa} a,
\end{equation}
where $\kappa$ is the cavity decay rate.
Similar to the three qubit case studied before, we assume there are no additional decay rates (e.g., off chip), other than to the waveguide or to the cavity, and so we assume $\kappa$ is the dominant decay process.

The cavity-emitted spectrum is obtained from
\begin{equation}
S_{\rm cav}(\omega)
= {\rm Re}
\int_0^\infty
\braket{a^\dagger(t_{\rm ss})a(t_{\rm ss}+\tau) }
e^{i(\omega-\omega_L)}
d\tau,
\end{equation}
where $t_{\rm ss}$ refers to steady state.
For all the driven JC calculations below, we have carefully checked for numerical convergence by increasing the photon number states to up to a maximum of $N=100$. 

\begin{figure} [h]
\centering
\includegraphics[width=0.95\columnwidth]{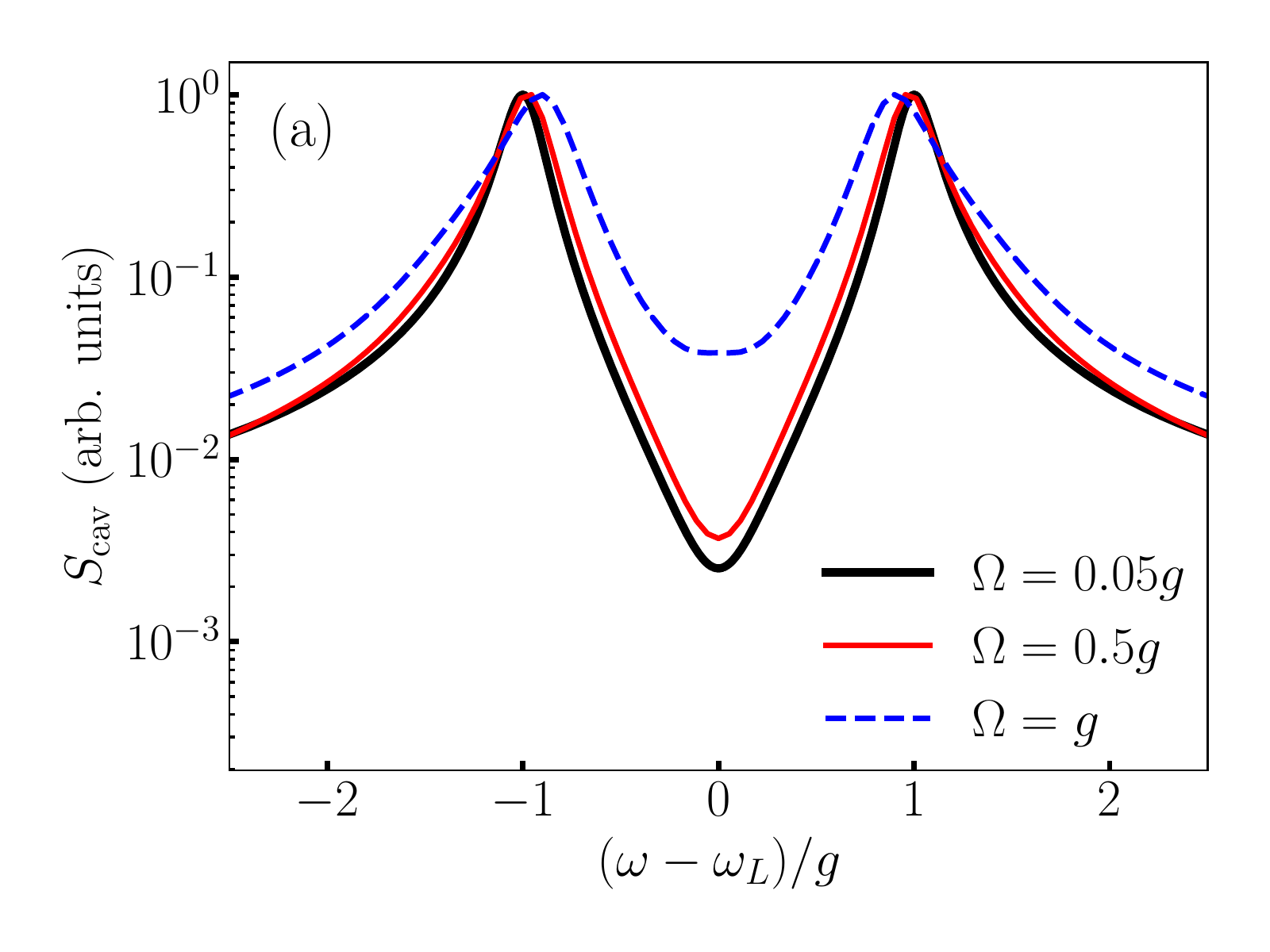} \\
\includegraphics[width=0.95\columnwidth]{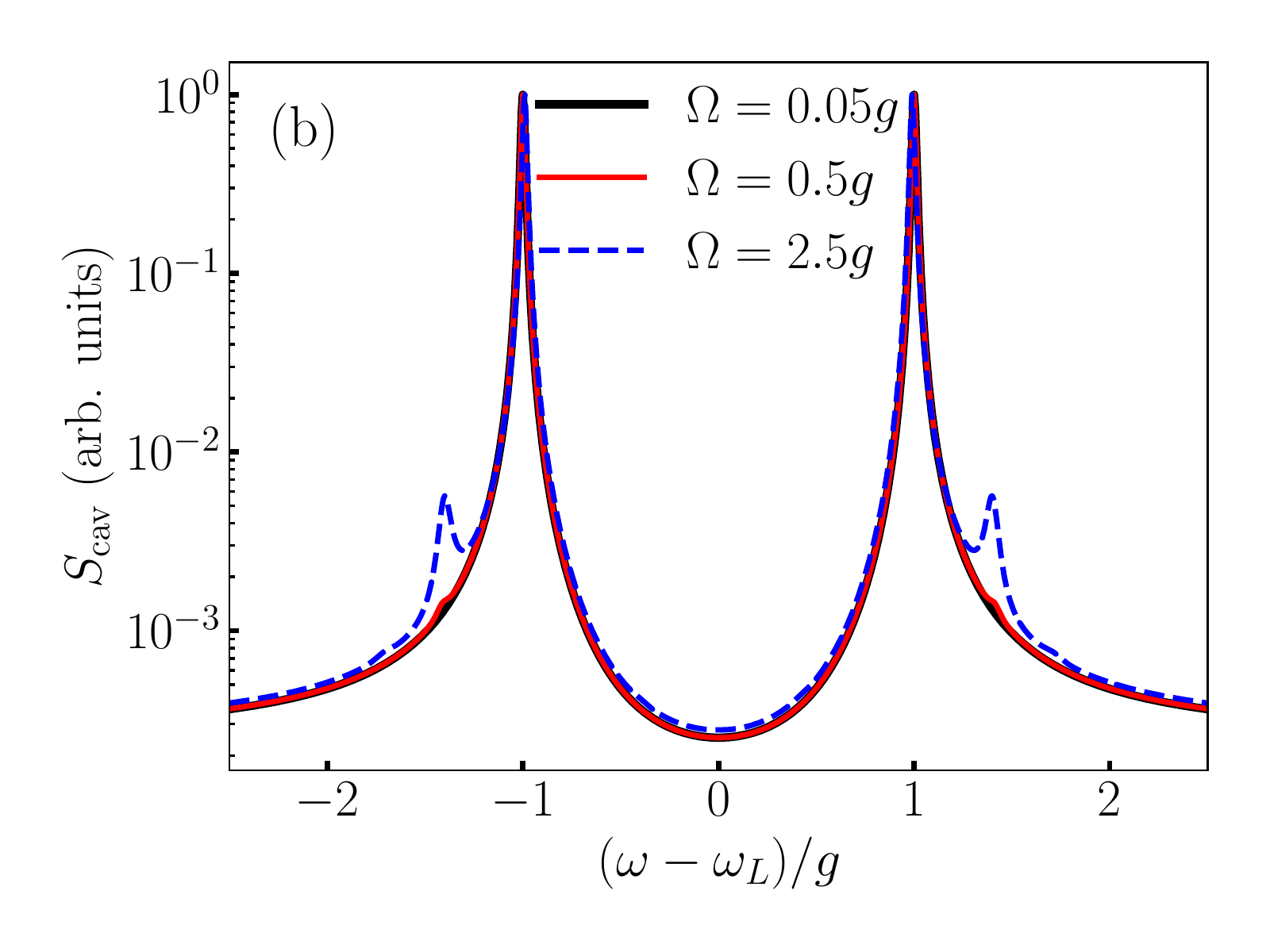} 
 \caption{Cavity-emitted spectra of the
 driven JC solution (single qubit-cavity system) for different qubit drive strengths [Eq.~\eqref{hJC}], with $\kappa=\gamma_p$ (which yields equivalent broadening in the linear strong coupling regime, in the Markov regime). (a) A driven system with $\kappa = \gamma_p$, to yield an equivalent linear polariton to the three qubit system. (b) Solution with a smaller value of the decay rate $\kappa = 0.1 \gamma_p $, where additional peaks can be partly observed for stronger pumps.  
 }
 \label{JC_Model}
\end{figure}

Figure \ref{JC_Model_en} shows the familiar JC energy-level ladder system for the dressed states, in the absence of optical pumping. Although the lower two polariton states
have an equivalence with the three qubit system in the Markov regime 
(namely, both can yield vacuum Rabi oscillations), this is only for linear excitation. Clearly the nonlinear states are quite different, and it is also well known  that accessing the higher-lying states of the JC system is notoriously difficult~\cite{Schuster2008,Kasprzak2010,Illes2015,Larson2021}, and typically one needs
very large $g/\kappa$ ratios, e.g., with state-of-the-art circuit QED  systems~\cite{Fink2008}, or/and very specialized spectroscopy techniques. This makes the observation and exploitation of the nonlinear JC states very challenging. Part of the problem in the JC system is that the dissipation also scales with the photon number, causing increasing dissipation for the higher lying excitations.

In order to compare our driven three qubit results with the (dissipative) JC solution, we choose the same value of $g$, and set $\kappa=\gamma_p$, since this yields similar polariton resonances with weak excitation [see Fig.~\ref{fig:roots}].
However, as can be seen in Fig.~\ref{JC_Model}(a), the observable peaks in a JC system are too broad to allow any discernible features from higher-order spectral peaks.
For clarity, we also show the solution with an order of magnitude reduction in the cavity decay rate, $\kappa=0.1 \gamma_p$; in this case, we can partly observe some of the anharmonic ladder states [Fig.~\ref{JC_Model}(b)], though  these are rather weak, even on a log scale.
 To be clear, in Fig.~\ref{JC_Model}(b), with suitable pumping, we barely observe a peak at $(\omega - \omega_L) = \sqrt{2}$ which corresponds to the second excitation manifold in Fig.~\ref{JC_Model_en}, and an even smaller peak  at  $(\omega - \omega_L) = \sqrt{3}$, corresponding to the third excitation manifold (or fourth energy level in Fig.~\ref{JC_Model_en}). This is significantly different from the results observed in Fig.~\ref{fig:energylev}, where for much weaker pump fields we can already see nonlinear spectral peaks representing different transitions between the dressed states of the three qubit system, and these are all clearly resolved even on a linear scale.

Note that in contrast to the dissipate JC system, in the three qubit-waveguide system there is no direct cavity decay---the full solution for photons and decay is automatically captured through the
couplings between the qubits and waveguide modes; the side qubits act as mirrors
and yield multiple cavity resonances, as discussed earlier, with the lower two polaritons having a decay rate of $\gamma_p/4$, which is analogous to a broadening of $\kappa/4$ only in the linear spectrum of a dissipative JC model~\cite{PhysRevB.60.13276}.


\subsection{Nonlinear Non-Markovian regime}

Next, we investigate the non-Markovian regime making use of the MPS approach seen in Sec.~\ref{subsection:MPS}. 
Here we fully account for the effects of retardation and nonlinearities on the pump-induced spectra.

In the nonlinear regime, the spectrum in the presence of retardation effects is shown in Fig.~\ref{fig:MPSresults}.  
We choose three example delay times (on-resonance) and plot the spectrum in terms of $g_{\rm eff}$ [Fig.~\ref{fig:MPSresults}(a)] and in terms of $g$ [Fig.~\ref{fig:MPSresults}(b)]. We can observe how the peaks, which can be explained through the dressed energy ladder states, do not depend on the retardation when considered in terms of $g_{\rm eff}$, although they appear at different positions when computed in terms of $g$.

In Figs.~\ref{fig:MPSresults}(c) and (d), the output field spectra are calculated for various delay times with an on-resonance pump [Fig.~\ref{fig:MPSresults}(c)], and an off-resonance one with $\Delta_n = g_{\rm eff}/2$ [Fig.~\ref{fig:MPSresults}(d)], where $\Omega = 0.5 \gamma_p$ in both cases. The spectral peaks seen in Fig.~\ref{qutip} are also observed in this regime. We recognize how these peaks do not depend on retardation  as they appear as straight vertical lines. However, the splitting observed on resonance in the Markovian regime disappears as the delay times increase.
Also, note that the retardation-induced peaks become narrower for increased delay times.
\begin{widetext}
    \begin{minipage}{\linewidth}
        \begin{figure}[H]
            \centering
     \includegraphics[width=0.45\columnwidth,trim={1cm 0 0 0},clip]{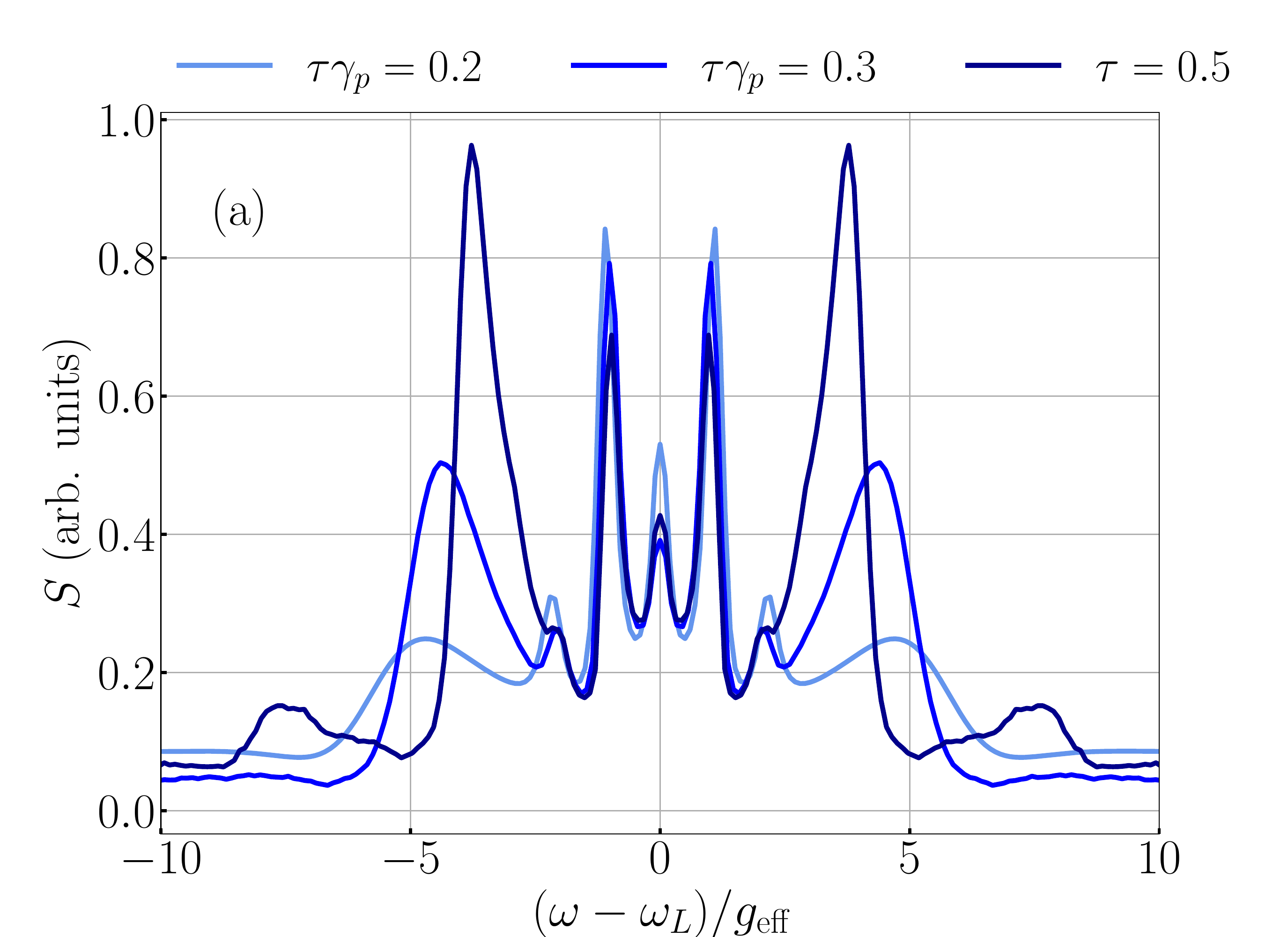} 
     \includegraphics[width=0.45\columnwidth,,trim={1cm 0 0 0},clip]{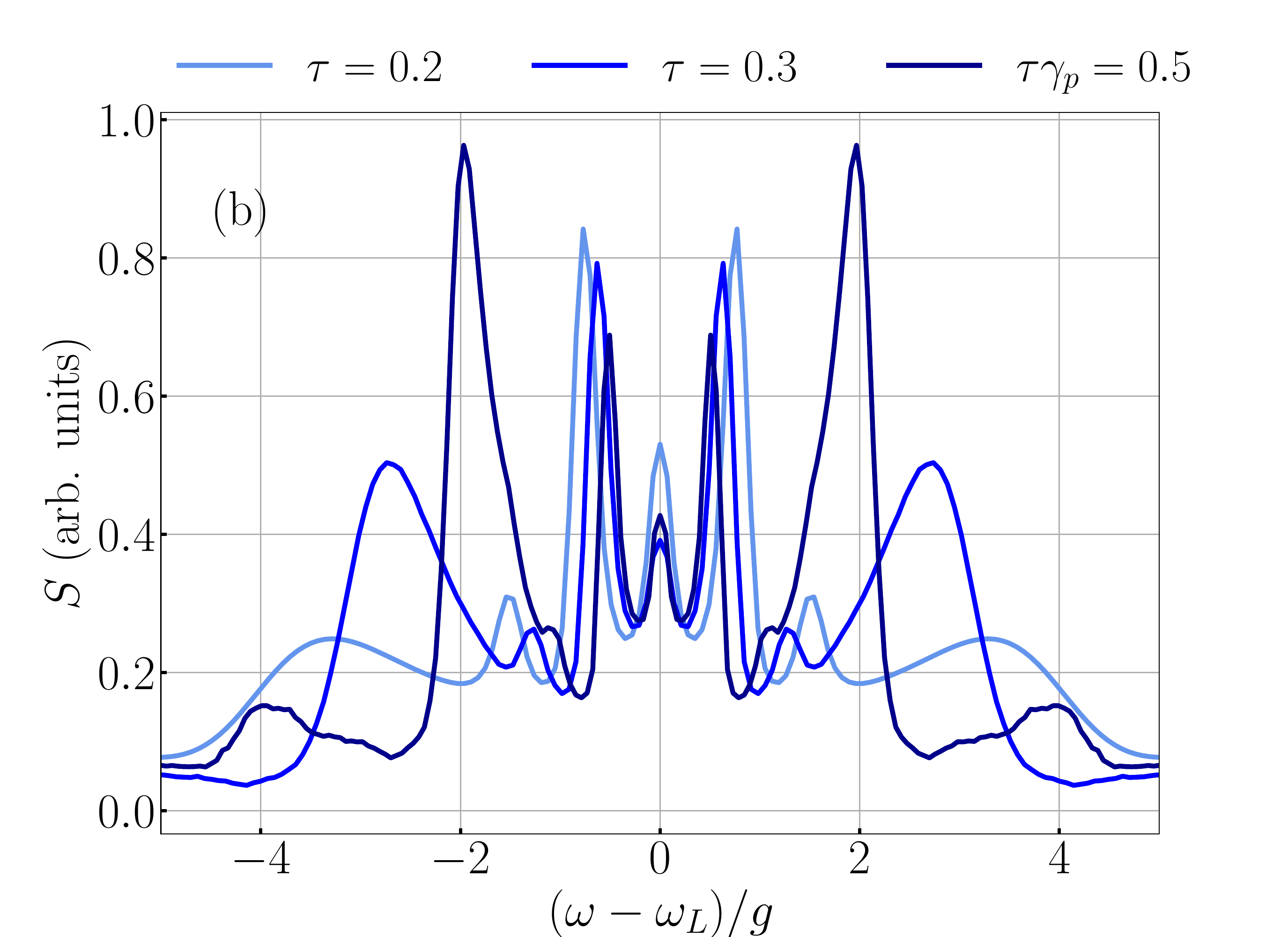}\\
     \includegraphics[width=0.49\columnwidth]{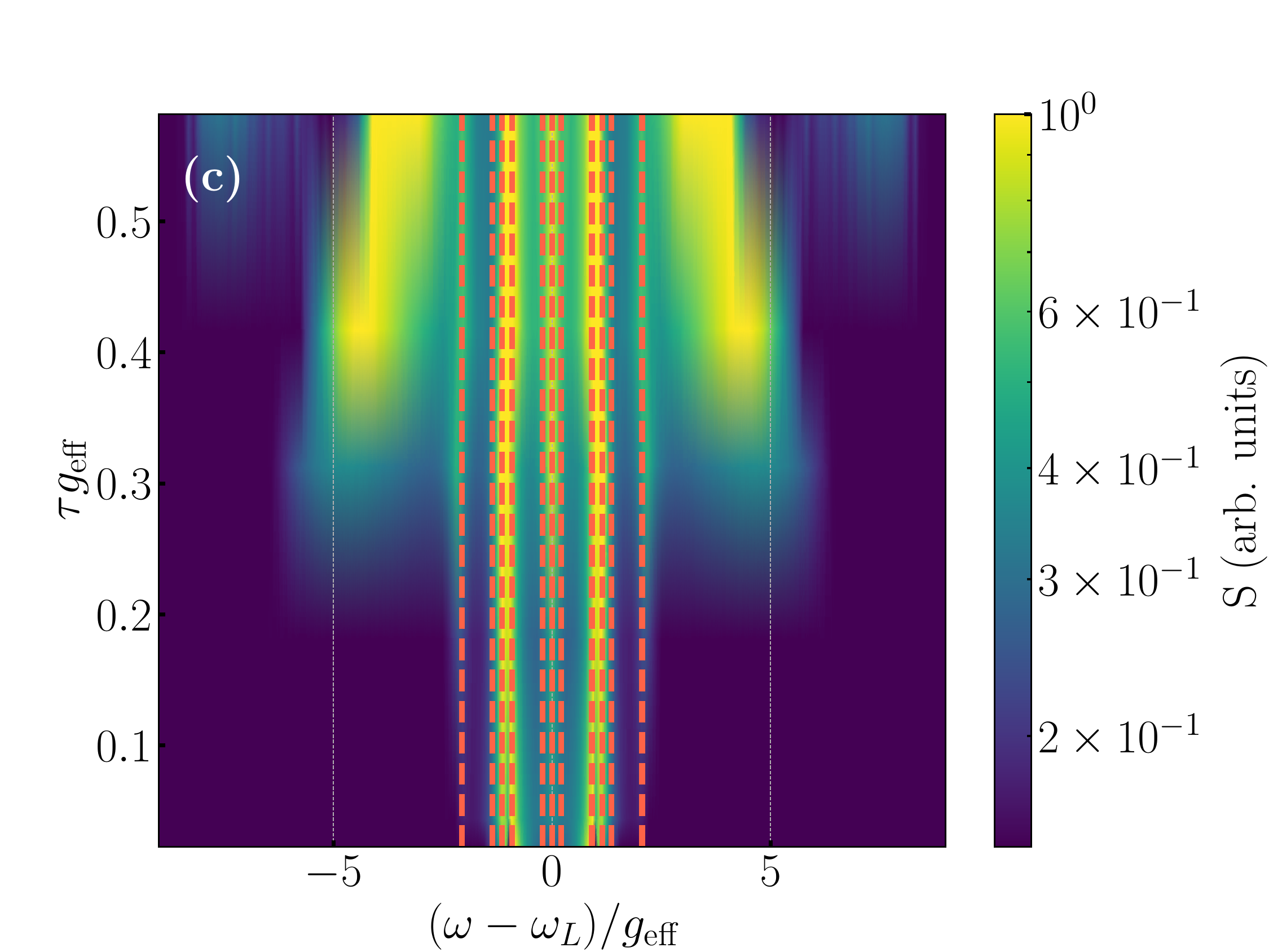} 
 \includegraphics[width=0.49\columnwidth]{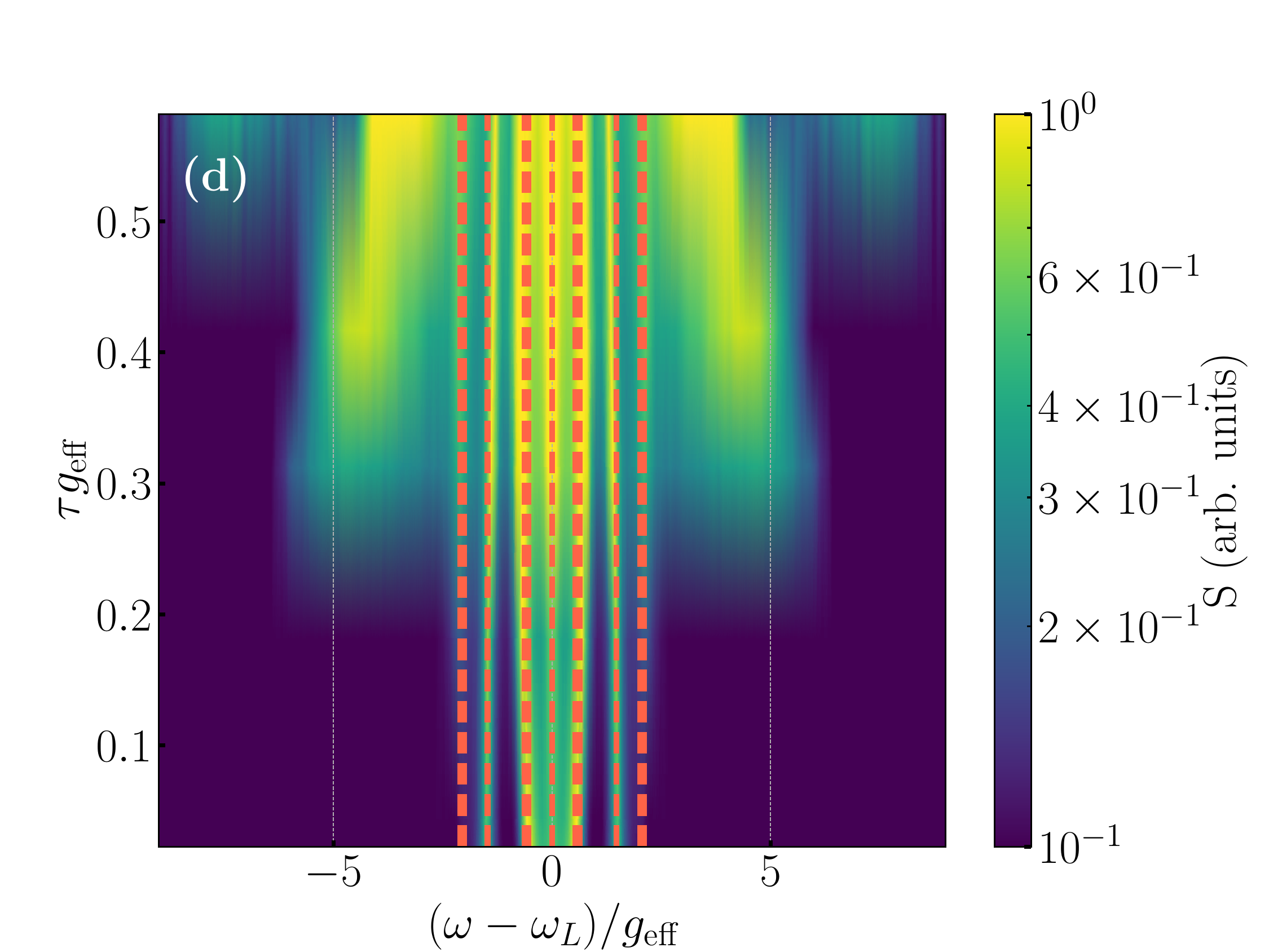} 
 \caption{ 
     Output spectra of a three-qubit system (with $\gamma_m / \gamma_p = 10$) calculated using MPS with the probe dot pumped with $\Omega=0.5 \gamma_p$. (a) and (b) show spectra with three delay times ($\tau \gamma_p = 0.2$, $\tau \gamma_p = 0.3$ and $\tau \gamma_p = 0.5$)  as a function of $(\omega - \omega_L)/g_{\rm eff}$ and as a function of $(\omega - \omega_L)/g$, respectively. Note the different frequency scales. (c) and (d) show contour plots with similar calculations of the output spectrum as a function of the delay times, with $\tau _{\rm max} = 0.5/ \gamma_p$. (c) On-resonance pumping. (d) Off-resonance with $\Delta_n = g_{\rm eff}/2$. Red dashed lines show the transition lines previously calculated in the Markovian regime.} 
 \label{fig:MPSresults}
        \end{figure}    
    \end{minipage}
\end{widetext}

In addition to the peaks identified from the Markovian dressed-states, there are now some extra peaks that emerge for sufficiently large  values of $\tau$ that cannot be explained in the Markovian limit, which also vary depending on the retardation (time delays). It is important to note that these additional resonances are not due to higher order cavity modes, since the free spectral range on this system (FSR), defined as $\Delta \omega_{FSR} \approx \frac{2 \pi}{\tau_{RT} \gamma_p}$ where $\tau_{RT}=2 \tau$ is outside the frequency regime shown. For instance, for $\tau \gamma_p=0.3$, $\Delta \omega_{FSR} = 10.47 \gamma_p$ which is outside of the spectral limits in Fig.~\ref{fig:MPSresults}. Thus the additional nonlinear peaks with retardation are not related to higher-order (cavitylike) modes.

\begin{figure}[hbt]
    \centering
    \includegraphics[width=0.9\columnwidth]{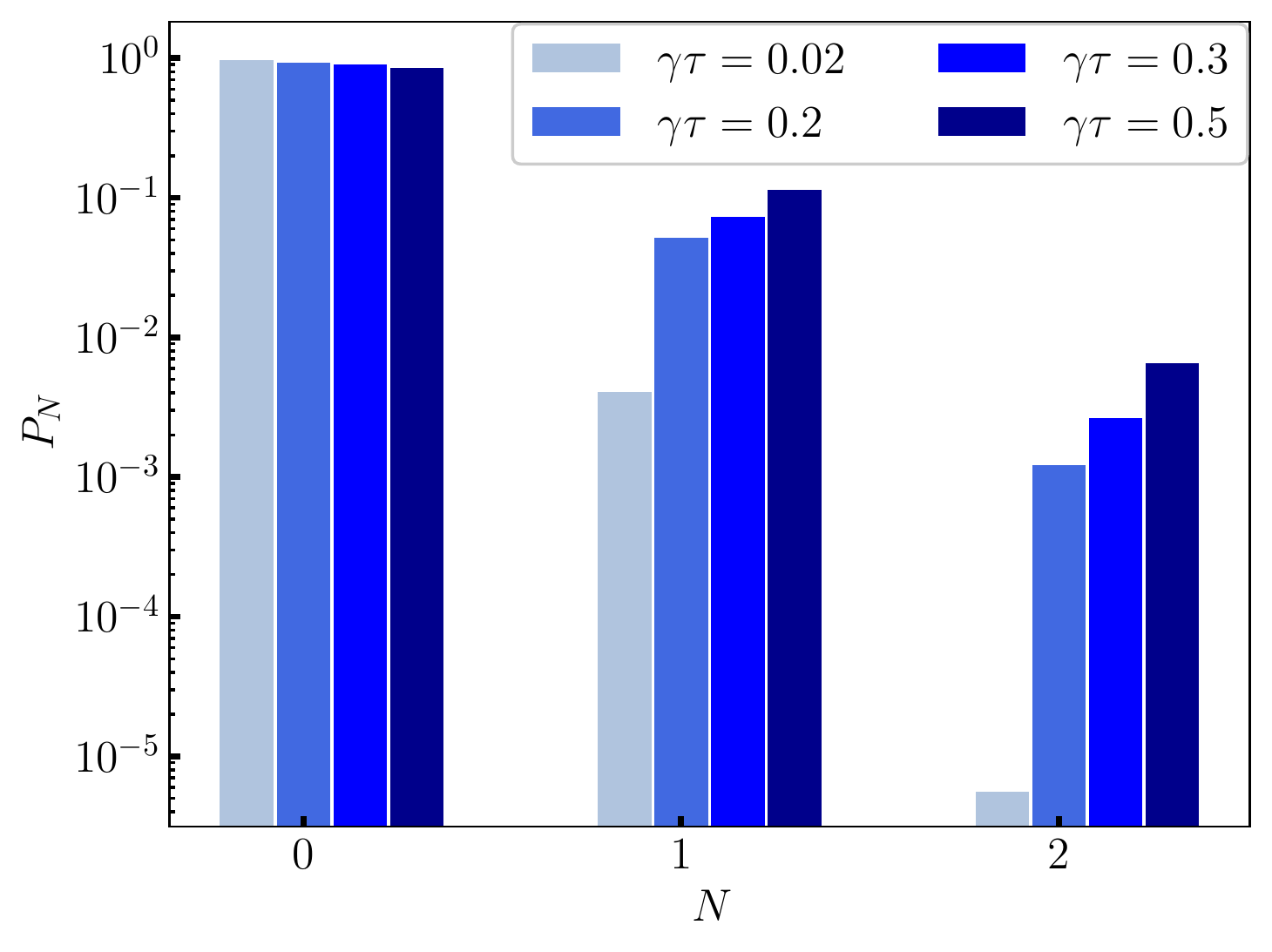}%
    \caption{Log scale of the photon probabilities for the first few photon states in the waveguide cavity region, for the case of resonant pumping (the case with off resonant pumping is similar, and thus is not shown).}
 \label{Pn}
  \end{figure}

The retardation-induced new peaks move in frequency and get more pronounced as the delay time increases, suggesting that this is a purely non-Markovian effect. To help us understand the origin of these new peaks, we also calculate the photon probabilities ($P_N$), 
by projecting the photon number operators on each time bin and calculating the probability of having zero or one photon ($\ket{0}\bra{0}$ and $\ket{1}\bra{1}$ respectively), and combining these results to get the probability of having zero ($P_0$), one ($P_1$) and two ($P_2$) photons in the part of the waveguide confined between the mirror qubits. We investigate this for four different retardation values to see how they vary as a function of $\tau$, for $\Omega=0.5 \gamma_p$ and $\Delta_n = 0$ (Fig.~\ref{Pn}). 

We observe that the probability of having two photons increases for higher values of $\tau$, although it stays considerably low in all case studies (for both on and off resonance calculations). In all our simulations, we also observe significant photon bunching in our system, which becomes more pronounced for longer time delays.

A similar phenomenon has been previously reported in Refs.~\cite{cajitas,PhysRevA.106.013714}, for time-delayed feedback, in the case of a single TLS in front of a mirror, where they found that  new non-Markovian resonances appear at frequencies $\omega = (\phi + 2\pi \mathbb{Z})/\tau$,
with $\mathbb{Z}$ the set of integers, $\phi$ the phase of the mirror and $\tau$ the retardation time between the mirror and the TLS. In our system, we do not have a single phase since there are three different phases involved due to the interaction between the three qubits, but we can observe an analogous trend in the dependence of these new peaks with respect to the delay times. In addition, our qubits act as nonlinear TLSs and also have a frequency dependent reflection, making the analysis more complicated than the single perfect mirror.

This supports the idea that the additional resonances we obtain with retardation are mediated through entangled photon-matter states not present in the Markovian master equation solution (where the waveguide modes are traced out).
Thus, it is essential here to have a method (such as MPS) that treats the waveguide
photons as part of the system, allowing for entangled photon matter states.


\section{Conclusions}
\label{sec:conclusions}

We have presented a theoretical study of an optically pumped three-qubit waveguide system where the side qubits act as mirrors, creating a cavitylike system with a probe qubit in its center.
We have theoretically modelled this system, starting with a linear model (for reference) and then accounting for various nonlinear interactions, where we first studied the Markovian limit by solving the medium-dependent (waveguide-qubit) master equation. The Markovian regime  allowed us to compare our results with a cavity system to establish a $\gamma_m / \gamma_p$ ratio that resembles the cavity behavior with the same characteristic Rabi splitting, and second, with the well-known JC model. With optical pumping, we observed new resonances including four resonances with a splitting near the one photon JC resonances, showing signatures of multi-quanta effects  beyond weak excitation. 

The nonlinearities were further explained by computing the dressed energy levels, where we first connected the natural dressed state basis (in the absence of any optical pumping) with the bare state basis. Then, we added  an optical pump field and computed again the dressed energy levels, as well as the dressed state populations, and subsequently the spectral peaks with their corresponding transitions. For comparison, we also showed the first few nonlinear states of a driven JC system, which was shown to yield a drastically different nonlinear response. Moreover, the higher order resonances of the driven JC system were not visible with a comparable level of dissipation.

Finally, we extended the model to include retardation and non-Markovian dynamics, solving the Hamiltonian with MPS, and comparing results for various values of time retardation. We observed how the resonances previously identified in the Markovian limit, do not depend on retardation {\it if} shown in terms of an effective coupling rate, $g_{\rm eff}$; we then found additional peaks that cannot be seen in the Markovian regime at all, which depend on the delay times; these peaks become narrower for higher delay times (which is a purely non-Markovian effect). Photon probabilities were also  calculated with MPS, showing low values of the two photon probabilities (although in an increasing trend with longer retardation times).
These new resonances stem from the excited qubit states, that cannot be seen in the Markovian regime. 

\medskip

\medskip

\acknowledgements
This work was supported by the Natural Sciences and Engineering Research Council of Canada (NSERC),
the National Research Council of Canada (NRC),
the Canadian Foundation for Innovation (CFI), and Queen's University, Canada.

\bibliography{refs_3dots}

\end{document}